\documentclass[prl,aps,superscriptaddress,reprint,twocolumn]{revtex4-2}
\usepackage{haimi}
\usepackage{graphicx}
\usepackage{xfrac}
\usepackage[normalem]{ulem}
\usepackage[caption=false]{subfig}
\usepackage{amsmath,amssymb}

\begin{document}

\title{Fate of Discrete Time Crystals in Finite-Size Cavity QED Systems}

\author{Haimi Nguyen}
\affiliation{Department of Chemistry, Columbia University, NY 10027, USA}
\author{David R.\ Reichman}
\affiliation{Department of Chemistry, Columbia University, NY 10027, USA}

\date{\today}

\begin{abstract}

It has been shown that Discrete Time Crystal (DTC) behavior may arise in the periodically driven, open Dicke model in the thermodynamic limit ($N \rightarrow \infty$).  In this work, we demonstrate that finite $N$ non-mean-field fluctuations in this model can lead to decay rates and amplitudes of transient DTCs that display a non-trivial non-monotonic behavior with an increasing number of atoms. We show that the second-order cumulant expansion fails to capture the finite-size dynamics, while a stochastic version of the Truncated Wigner Approximation (TWA) is remarkably accurate, particularly in the large atom ensemble limit. Using the TWA and the Monte Carlo Wave Function (MCWF) approach, we identify three distinct dynamical regimes of transient DTC behavior in the finite-size Dicke model.  We outline the mechanism behind the non-monotonic finite size dynamical decay of DTC order and discuss the implications for experiments performed on finite-sized atomic assemblies.
\end{abstract}

\maketitle

Discrete Time Crystals (DTCs) have emerged as a fascinating non-equilibrium phase of matter, characterized by robust subharmonic oscillations that break discrete time-translation symmetry. A primary challenge in maintaining this phase is ``overheating," where the periodic drive erases temporal order by pumping energy into the system. To counteract this, systems may be designed to leverage quantum stabilization induced by disorder or utilize an external bath to dissipate energy.  Transiently stable DTCs have been realized in experimental settings such as trapped atomic ions~\cite{Zhang2017}, diamond vacancy centers~\cite{Choi2017}, superfluid quantum gases~\cite{Smits2018}, and driven atom-cavity systems~\cite{Keler2021,Taheri2022}. 

The Dicke model, originally introduced to describe the collective interaction between an ensemble of $N$ indistinguishable two-state atoms and a bosonic cavity mode~\cite{Hepp}, has emerged as a platform to explore emergent phenomena in a quantum optics setting. In particular, the open Dicke model exhibits complex dynamics and rich non-trivial phase transitions~\cite{Kirton2018}. In 2018, Gong \textit{et al.}~\cite{Gong2018} showed that DTC behavior can arise in a periodically driven open Dicke model with cavity loss. It may be argued that due to the permutation symmetry of the atoms, the DTC behavior exhibited by this model arises from effective few-body physics and is not an example of non-trivial DTC stabilized by many-body interactions~\cite{yao}.  We do not draw such distinctions in this work, simply focusing on the model's non-trivial dynamical behavior. The stability of this phase discovered by Gong \textit{et al.} is, however, only guaranteed in the mean-field limit where $N=\infty$. Experimental realizations in platforms such as trapped ions, superconducting qubits, and atom-cavity systems are inherently finite sized. A critical open question is how finite-size effects influence the stability and lifetime of the DTC phase. Gong \textit{et al.}~\cite{Gong2018} predicted that the decay rate of the DTC would decrease monotonically with $N$, and suggested that the second-order cumulant expansion might be a viable approach to study finite-size fluctuations in cavity QED-based DTC systems.

In this work, we discuss the behavior of a finite number of identical two-level atoms coupled to a lossy cavity mode with a periodic time-varying coupling.  Via exact numerical simulations we show that in fact the amplitude and stability of DTC behavior varies in complex and non-monotonic manner with increasing $N$.  In addition, in some regimes DTC behavior remains completely stable over many oscillation periods.  We demonstrate that the second-order cumulant approach is not a reliable method for describing finite-size effects, but a variant of the truncated Wigner approximation (TWA) appears to be remarkably accurate for reproducing the dynamical behavior seen in both the driven and undriven versions of the Dicke model.  This fact, useful in its own right for describing cavity QED systems, enables the efficient description of transient DTC behavior for long times in systems far larger than can be approached by exact methods.

We consider an ensemble of $N$ indistinguishable two-level atoms coupled to a single cavity mode, described by the Hamiltonian ($\hbar=1$):
\begin{equation} \label{eqn:hamiltonian}
    H(t) = \omega_c \hat{a}^\dagger \hat{a} + \omega J_z + \frac{2\lambda(t)}{\sqrt{N}} (\hat{a}^\dagger + \hat{a}) J_x,
\end{equation}
where $J_\gamma = \sum_{i=1}^N \frac{1}{2}\sigma_\gamma^{(i)}$ are collective spin operators, $\sigma_\gamma^{(i)}$ ($\gamma=x,z)$ are Pauli matrices representing Hamiltonian terms of individual two-level atoms, and $\hat{a}^\dagger$ is the bosonic creation operator for the cavity mode. Here, $\omega_c$ and $\omega$ denote the cavity and atomic frequencies, respectively. We define the detuning parameter as $\epsilon = (\omega - \omega_c)/2$. In the standard Dicke model, the coupling between the atoms and the cavity is constant, $\lambda(t) = \lambda_0$. For the modulated model, the coupling $\lambda(t)$ periodically varies as $\lambda(t) = \lambda_0$ for $0 \le t~(\bmod~T) < T/2$, and $\lambda(t) = 0$ for $T/2 \le t~(\bmod~T) < T$ with $T=2\pi$. The system evolves under the Lindblad master equation with cavity loss rate $\kappa$:
\begin{equation}
    \dot{\rho} = -i[H(t), \rho] + \kappa \left( \hat{a}\rho \hat{a}^\dagger - \frac{1}{2}\{\hat{a}^\dagger \hat{a}, \rho\} \right).
\end{equation}
Because the cavity loss is collective, the atoms remain indistinguishable, enabling the use of the collective representation $\sigma_\alpha=\frac{2}{N}J_\alpha$ for all simulations.
In the thermodynamic limit ($N=\infty$) of the undriven model, the system undergoes a transition to a superradiant state characterized by spontaneous symmetry breaking associated with non-zero values of $\langle \hat{a}\rangle,~\langle \sigma_x \rangle$, and $\langle \sigma_y \rangle$ when $\lambda_0$ exceeds the critical value of $\frac{1}{2}\sqrt{\frac{\omega}{\omega_c}\left(\omega_c^2+\kappa^2/4\right)}$~\cite{Kirton2018}.  In this Letter, we will quantify the DTC order using the expectation value $\langle \sigma_x \rangle$ unless specified otherwise.  Finally, we mention here that for a realistic cavity QED model, gauge invariance requires the addition of a diamagnetic term to the Hamiltonian, which destroys non-trivial emergent phase behavior~\cite{Rzazewski}. However, models without this term exist in other settings, e.g. in solid-state magnonic systems~\cite{Kono}, and we thus proceed without modification to study the model specified above. 

To explore the finite-size system dynamics, we employ three complementary numerical approaches.

For very small system sizes ($N \le 10$), we solve the Lindblad master equation exactly by propagating the total density matrix through the QuTiP package's \textit{mesolve} routine~\cite{Johansson2012,Johansson2013}. Due to the global nature of the Lindblad operator employed in this model, the atom ensemble is represented in the collective spin basis, significantly reducing the computational demand. 
However, the finite photon basis for the cavity must be expanded rapidly for convergence as $N$ increases.
To work around this expansion, for $10 < N \le 20$, we utilize the Monte Carlo Wave Function (MCWF) quantum jump method, which unravels the density matrix evolution by stochastic quantum trajectories of wavefunctions~\cite{Dalibard1992}. We average over 5,000 trajectories to ensure convergence of the observables.

For large $N$, we employ a recently developed form of the TWA~\cite{Hosseinabadi2025}. This semi-classical method approximates the quantum dynamics by evolving an ensemble of classical trajectories initially sampled from the Wigner distribution of the quantum state. The equations of motion are:
\begin{align}
    \dot{\alpha} &= -i\omega_c \alpha - \frac{i\lambda(t)}{\sqrt{N}} \sum_i s_x^{(i)} - \frac{\kappa}{2}\alpha - \frac{\xi_\kappa}{2},\label{eqn:TWA0} \\
    \dot{s}_x^{(i)} &= -\omega s_y^{(i)},\label{eqn:TWA1} \\
    \dot{s}_y^{(i)} &= \omega s_x^{(i)} - \frac{4\lambda(t)}{\sqrt{N}} \Re[\alpha] s_z^{(i)}, \label{eqn:TWA2}\\
    \dot{s}_z^{(i)} &= \frac{4\lambda(t)}{\sqrt{N}} \Re[\alpha] s_y^{(i)},\label{eqn:TWA3}
\end{align}
where $\xi_\kappa$ is a complex Gaussian noise term centered at 0 satisfying $\overline{ \xi_\kappa(t) \xi_\kappa^*(t') } = 2\kappa \delta(t-t')$. 
Here, the photon field is initialized by sampling complex amplitudes $\alpha$ from the Gaussian Wigner function of the bosonic coherent state $|\alpha_0\rangle$, $W(\alpha)=\frac{2}{\pi}\exp{\left[-2\left|\alpha-\alpha_0\right|^2\right]}$. This distribution is centered at the coherent state's complex amplitude with a standard deviation of $\frac{1}{2}$ to account for quantum vacuum fluctuations. 
For the spin degrees of freedom we utilize a discrete sampling scheme. Each spin $s^{(i)}$ is initially oriented along the $z$-axis with transverse components $s^{(i)}_x$ and $s^{(i)}_y$ randomly assigned values of $\pm 1$, before being rotated to the desired initial state.
A key technical advantage of this formulation in the open Dicke model is that the computational cost of TWA remains \textit{independent of the atom number $N$}. Because the photon noise $\xi_\kappa$ is the only source of stochasticity during propagation, we do not need to track $N$ individual trajectories. Instead, 
we perform random sampling on the discrete initial $(s_x, s_y)$ pairs for all trajectories and tally the occurrences of each state. Then, we computationally evolve only four trajectories representing the four unique initial states but take into account their trajectory multiplicities in our collective calculations such as the total spin sum.
This allows us to efficiently simulate large ensembles of any size. However, the accuracy of TWA is still to be tested.

We compare these results with the Second-Order Cumulant Expansion (CE), a standard approach for including fluctations beyond the mean-field level in collective spin models~\cite{Kerber}. This method truncates the hierarchy of correlations by setting all cumulants of order three and higher to zero, effectively assuming a Gaussian state at all times. In this work, we distinguish between two implementations: the Partial Second-Order Cumulant Expansion (PSC) used in literature~\cite{Kirton2017} to find steady-state values and the Full Second-Order Cumulant Expansion (FSC).  The PSC removes terms that break the $\mathbb{Z}_2$ symmetry of the undriven model. Because these eliminated variables are the very order parameters required to characterize a DTC, the PSC is fundamentally unsuitable for simulating DTC dynamics. In contrast, our FSC implementation restores these terms. The FSC equations of motion can be found in the Supplementary Materials (SM).

\begin{figure}
\includegraphics[width=\linewidth]{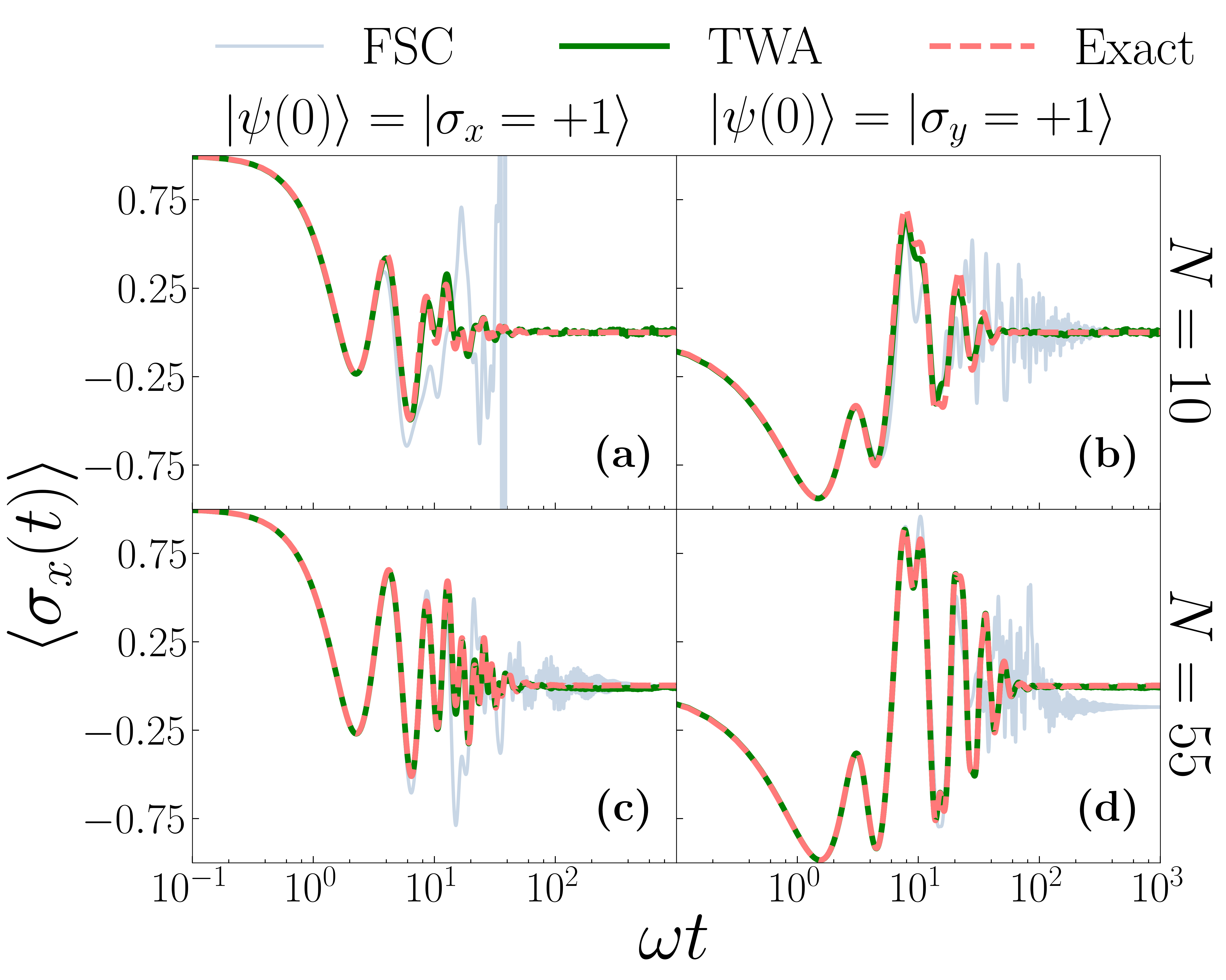}
\caption{
Dynamics of magnetization $\langle \sigma_x(t) \rangle$ in the standard undriven Dicke model. Results are compared against results from exact methods (red dashed), Truncated Wigner Approximation (TWA) (green solid), and Full Second-order Cumulants (FSC) (grey solid). The grid compares different atom ensemble sizes, $N=10$ (upper row) and $N=55$ (lower row), and initial spin orientations, positive $x$-axis (left column) and positive $y$-axis (right column). The cavity is initialized in the vacuum state. Parameters: $\epsilon=0.0$, $\kappa=0.05$, and $g=0.6$.}
\label{fig:method_comparison}
\end{figure}

\begin{figure}
    %\centering
    \includegraphics[width=\linewidth]{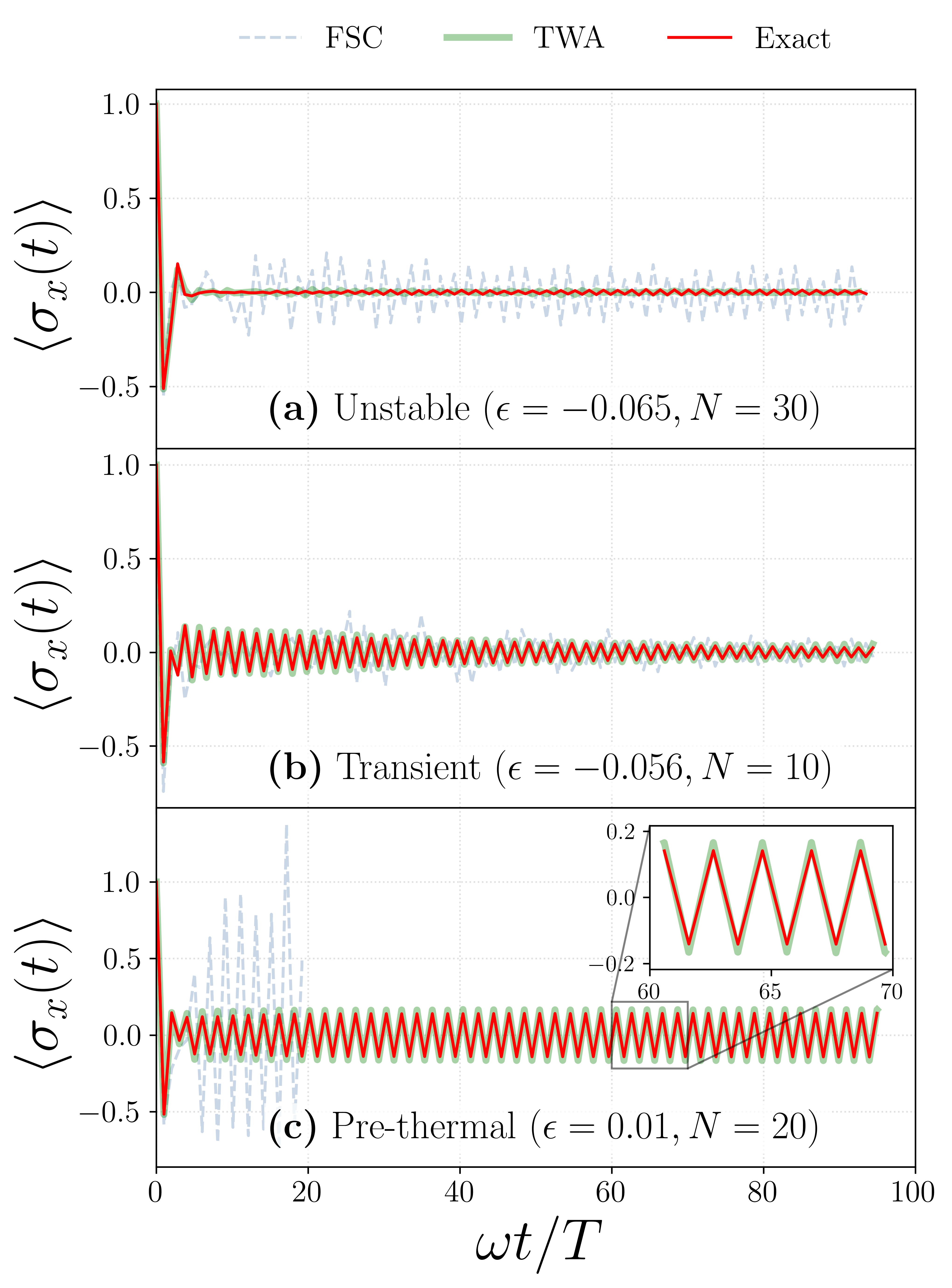}
    \caption{
    Stroboscopic dynamics of magnetization $\langle \sigma_x(nT) \rangle$ in the periodically modulated Dicke model. Results for exact methods (red solid), Truncated Wigner Approximation (TWA) (green solid), and Full Second-order Cumulant (FSC) (grey dashed). The initial state consists of the cavity in a vacuum state with the with the collective spin state oriented along the positive $x$-axis. Three dynamical regimes are identified: \textbf{(a)} Unstable DTC, characterized by rapid decay of temporal order ($\epsilon=-0.065,~N=30$); \textbf{(b)} Transient DTC, where order persists for hundreds of cycles but clear decay is evident ($\epsilon=-0.056,~N=10$); and \textbf{(c)} Pre-thermal DTC, where order remains essentially stable in period and amplitude over the entire simulation timescale ($\epsilon=0.01,~N=20$). The inset in \textbf{(c)} provides a magnified view, highlighting the high degree of overlap between the TWA and exact results. Other parameters: $\kappa=0.05, \lambda_0=1.0$.}
    \label{fig:dtc_regimes_dynamics}
\end{figure}

Even though the FSC is a widely used tool for collective spin models, we find that it is unreliable for the problems of interest in this work as shown in Fig.~\ref{fig:method_comparison} and Fig.~\ref{fig:dtc_regimes_dynamics} for the standard and periodically modulated open Dicke models, respectively. While the system's density matrix naturally evolves into a non-Gaussian distribution through its highly coupled, non-linear equations of motion, the FSC constrains it to an essentially Gaussian form. 
For a strongly-coupled finite-size Dicke system, Gaussian statistics cannot capture fluctuations that span two symmetric manifolds, as expected when transient symmetry breaking occurs. 
On the other hand, the PSC, as detailed by Kirton \textit{et al.}~\cite{Kirton2017}, exhibits greater numerical stability precisely because it eliminates all the $\mathbb{Z}_2$ symmetry-breaking terms, but this comes at the expense of the ability to describe DTC-type dynamics.  As shown in the SM, when numerical stability is present, the FSC and the PSC yield accurate steady-state results for symmetric variables such as the cavity photon number. However, when it comes to symmetry-breaking order parameters (e.g. $\langle \sigma_x \rangle$), the FSC often fails to converge or produces inaccurate steady-state results, as shown in Fig.~\ref{fig:method_comparison}(a,d). 

In light of the breakdown of the FSC and PSC, we turn to the stochastic TWA approach. This approach has
the ability to describe non-Gaussian statistics through an ensemble of independent trajectories.

\begin{figure}
    \centering
    \includegraphics[width=\linewidth]{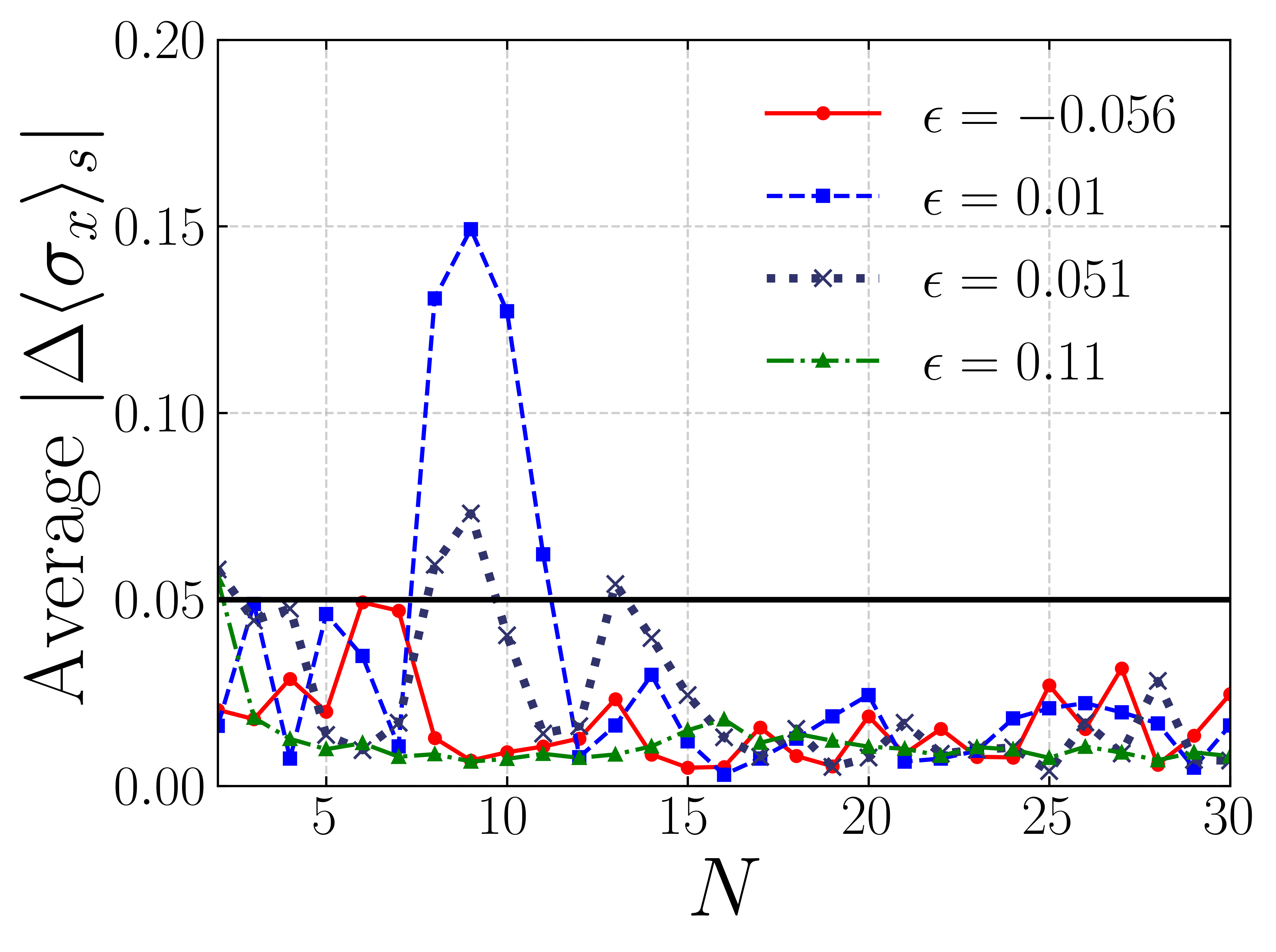}
    \caption{ 
    Average absolute difference in stroboscopic magnetization $\langle \sigma_x(nT) \rangle$ between exact methods and the Truncated Wigner Approximation (TWA) in a periodically modulated Dicke model. The difference is plotted against atom ensemble size $N$ for detuning parameters $\epsilon = -0.056$ (red solid, circles), $0.01$ (blue dashed, squares), $0.051$ (dark blue dashed, crosses), and $0.11$ (green dash-dotted, triangles). The horizontal solid black line represents the threshold below which the signal-to-noise ratio is insufficient to identify DTC behavior. Parameters: $\kappa=0.05, \lambda_0=1.0$.}
    \label{fig:error_plot}
\end{figure}

Surprisingly, despite its approximate semiclassical nature, the TWA performs remarkably well even in small-$N$ regimes (Fig.~\ref{fig:method_comparison},~\ref{fig:dtc_regimes_dynamics},~\ref{fig:error_plot}). We attribute this success in part to the open nature of the system: dissipation from photon loss rapidly destroys quantum coherence between the transiently symmetry-broken states. This decoherence suppresses superposition states in favor of statistical mixtures, rendering higher-order quantum corrections negligible.  The surprising near-quantitative accuracy of the TWA approach for the description of dynamics in cavity QED problems is of broad utility.  Here, we leverage this accuracy to describe in detail the fate of DTC behavior for finite $N$ in the model of Eq.~\ref{eqn:hamiltonian}.

The interplay between quantal and classical effects on distinct timescales appears to give rise to three distinct dynamical regimes for the order parameter $\langle \sigma_x(2nT) \rangle$:
\begin{enumerate}
    \item \textbf{Unstable DTC}: In regions of large frequency detuning and weak dissipation, the DTC order is either absent or decays to below 0.05 in amplitude after stabilization. This is depicted in Fig.~\ref{fig:dtc_regimes_dynamics}(a).
    \item \textbf{Transient DTC}: A DTC is observed but decays within timescales relevant to simulations and experiments. This is shown in Fig.~\ref{fig:dtc_regimes_dynamics}(b).
    \item \textbf{Pre-thermal DTC}: The system exhibits robust period-doubling oscillations with a vanishing decay rate on the long timescales we simulate. We conservatively call such behavior ``pre-thermal." The true long-time stability of such a phase is unclear and may be robust.  This is illustrated in Fig.~\ref{fig:dtc_regimes_dynamics}(c).
\end{enumerate}
The distinction between the transient and pre-thermal behavior may appear to be somewhat arbitrary; however, there is a marked empirical distinction between the decay envelopes in these two regimes, as can be seen in~Fig.~\ref{fig:dtc_regimes_dynamics}.

\begin{figure*}
    \centering
    \includegraphics[width=\textwidth]{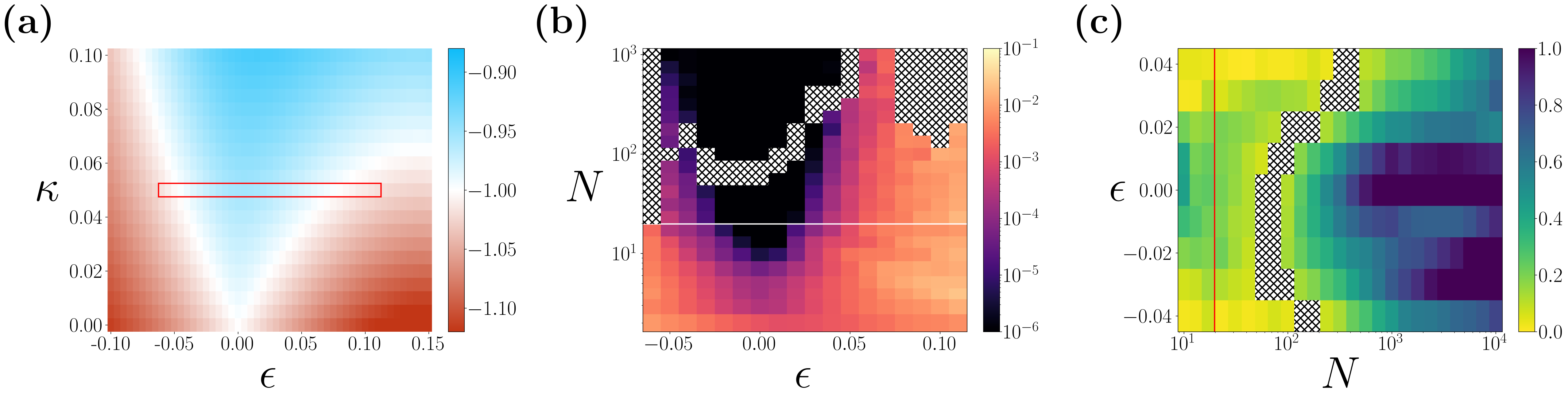}
    \caption{Global Discrete Time Crystal (DTC) stability and finite-size scaling via exact methods and the Truncated Wigner Approximation (TWA). \textbf{(a)} Stability phase diagram showing the magnitude of the largest Floquet multiplier (from linearized analysis) in the $\epsilon$-$\kappa$ plane. The red region denotes chaotic behavior, while the blue region shows stability.  The red box marks the parameter regimes in \textbf{(b)}. \textbf{(b)} Decay rate of the DTC order parameter $\langle \sigma_x(2nT) \rangle$ as a function of atomic ensemble size $N$ and frequency detuning parameter $\epsilon$. The white horizontal line indicates the transition from exact methods ($N\leq20$) to TWA ($N > 20$). \textbf{(c)} Ratio of the finite-size DTC amplitude of $\langle \sigma_x(2nT) \rangle$ to the mean-field value. The red vertical line separates values computed by exact methods ($N\leq20$) and TWA ($N > 20$). Hatch regions in \textbf{(b)} and \textbf{(c)} indicate regimes where the DTC phase is either absent or its amplitude decays below 0.05, the threshold shown in Fig.~\ref{fig:error_plot}. For all the heatmaps, the coupling strength is $\lambda_0=1.0$ with a vacuum cavity and spin polarized along the positive $x$-axis as the initial state.}
    \label{fig:combined_heatmap_figure}
\end{figure*}

With the DTC behaviors clearly defined, we turn to assessing the accuracy of TWA against exact methods in DTC-present regimes, as shown in Fig.~\ref{fig:error_plot}. 
Here, we benchmark the average absolute difference in the stroboscopic $\langle \sigma_x \rangle$ over 100 cycles across a range of frequency detuning parameters $\epsilon$ that exhibits both transient and pre-thermal DTC behaviors. 
In these regimes, TWA performs remarkably well when $N$ is larger than 20, consistently remaining within the 0.05 absolute error in $\langle \sigma_x \rangle$ (or 5\% error rate) indicated by the horizontal black line.
However, the agreement is not perfect for small $N$, particularly at $N=9$ for $\epsilon \in \{0.01,0.051\}$, where the DTC behavior is pre-thermal, and TWA predicts the correct DTC type but not the exact amplitude.
This benchmark provides confidence in using TWA for $N >20 $ in Fig.~\ref{fig:combined_heatmap_figure}(b,c) and justifies setting a 5\% error threshold, below which the signal-to-noise ratio is too low for accurate DTC evaluation.

Fig.~\ref{fig:combined_heatmap_figure}(a) surveys DTC stability across a broad range of parameters, which is quantified by the largest Floquet multiplier of the linearized, deterministic part of the TWA equations (see SM for details). 
This is equivalent to the linearized mean-field ($N=\infty$) equations, yielding an $N$-independent stability metric. The heatmap verifies a crucial claim by Gong \textit{et al.}~\cite{Gong2018} that cavity loss stabilizes the DTC order. Dissipation here plays a dual role: it acts as a cooling mechanism that drains energy injected by the drive and helps preserve DTC order, while it also acts as a damping source that eventually counteracts DTC stability. In Fig.~\ref{fig:combined_heatmap_figure}(a), the stabilizing ``cooling" effect dominates, as evidenced by the expansion of the stable DTC region (blue) with increasing cavity loss rate $\kappa$.

In contrast, finite-$N$ scaling displays rich behavior, as observed in Fig.~\ref{fig:combined_heatmap_figure}(b,c). 
To describe the distinction between mean-field and finite-$N$ behavior, we need to define stability metrics that quantify the fate of DTCs at finite $N$. 
In this Letter, for transient DTC regime, we select the decay rate of the DTC amplitude to be that metric. 
In pre-thermal regimes where the decay rate is vanishingly small, we select the ratio of the finite-$N$ to mean-field DTC amplitude as the metric.  
These measures are natural since both temporal decay and amplitude suppression are associated with the destruction of DTC order away from the mean-field limit.
We utilize exact methods for $N \le 20$ with a maximum simulation time of 1500 periods and TWA for $N > 20$ with a maximum simulation time of 5000 periods. Starting from the mean-field behavior shown in Fig.~\ref{fig:combined_heatmap_figure}(a), we isolate a region predicted to display DTC behavior with a transition to chaos at the periphery ($\epsilon \in [-0.06, 0.11]$ at $\kappa=0.05$, highlighted by the red box).
We then compute the corresponding DTC decay rates in Fig.~\ref{fig:combined_heatmap_figure}(b). 
Finally, we explore the regimes of $\epsilon \in [-0.04, 0.04]$, where pre-thermal DTC dominates, and compute the amplitude ratios as shown in Fig.~\ref{fig:combined_heatmap_figure}(c) for a wider range of $N$.

The non-monotonic stability of DTC order illustrated in Fig.~\ref{fig:combined_heatmap_figure}(b,c) presents a surprising departure from the intuition that signatures of DTC order should become uniformly more prominent as $N$ increases. It is natural to hypothesize, as in Ref.~\cite{Gong2018}, that the system asymptotically approaches the mean-field limit as the tunneling barrier required for thermalization grows with $N$. However, the results in Fig.~\ref{fig:combined_heatmap_figure}(b,c) show that in both pre-thermal and transient regimes, there is an intermediate region which exhibits a dip in DTC stability with increasing $N$. 
In transient DTC regimes, this appears as the light orange island observed for $N < 20$ and $\epsilon \in [0.05,0.11]$ in Fig.~\ref{fig:combined_heatmap_figure}(b). In pre-thermal regimes, it manifests in the hatched wall in Fig.~\ref{fig:combined_heatmap_figure}(c), where the DTC amplitude drops below 0.05 at $N \sim100-500$ before recovering and asymptotically approaching the expected behavior. 

Here we employ intuition afforded by the semi-classical nature of the TWA approximation. From the symmetry of the Hamiltonian in Eq.~\ref{eqn:hamiltonian}, there exist two degenerate Floquet attractors where all $\mathbb{Z}_2$ symmetry-breaking variables differ only by their signs. 
We label each attractor by how these variables relate to their mean-field counterparts: if they share the same sign, they are labeled as the in-phase attractor; otherwise, they are the out-of-phase attractor. 
To facilitate the understanding of the picture described next, we provide an HTML file for simulating the TWA trajectories for any $N$ and detuning, along with key screenshots of generated trajectories in the SM.

As shown in the SM, an intricate foliated boundary in parameter space separates the two stroboscopic mean-field attractors as defined above. The initial cloud of phase points, whose spatial extent decreases algebraically as $N$ increases, can straddle the boundaries of these two attractors.  In regimes of detuning parameters where, for example, pre-thermal behavior is seen, the cloud of phase points mostly localizes in a single effective attractor as $N \rightarrow \infty$. For finite $N$, the interplay between the size of the region in phase space occupied by initial state points, and the length scales associated with the irregular boundary between long-time attractors, yields the approach to the $N \rightarrow \infty$ non-monotonic in the amplitude of the observed oscillations.  A similar competition can be seen for detunings that place the system in the transient DTC regime.  In addition,
for transient regimes where $\epsilon$ deviates significantly from zero, the mismatch between the angular velocities of the spin and the cavity field scrambles the trajectories after each period, causing their amplitudes to cancel out upon averaging, leading to the non-monotonic scaling of the DTC decay rate. 

An important feature of the dynamics of our system is that the sum of the system's Lyapunov exponents that arise in the dynamical system defined by Eq.~\ref{eqn:TWA0}--\ref{eqn:TWA3} is equal and opposite to the cavity loss rate ($-\kappa$).
This constant phase-space contraction ensures that the system's dynamics is confined to strange attractors, periodic limit cycles, or fixed points.

For pre-thermal DTC regimes, the largest Lyapunov exponent is initially positive before quickly decaying to zero, showing a rapid transition from behavior described by strange attractors to that of a stable limit cycle.
In contrast, for transient DTC regimes, the largest Lyapunov exponent remains positive for significantly longer times, indicating a higher degree of dynamical unpredictability.

Current experimental platforms span the full range of $N$ in the phase behavior shown in Fig.~\ref{fig:combined_heatmap_figure}. Experiments operating with $N \sim 10-50$, commonly found in trapped ions and circuit QED, are firmly in the regime where the stability of the DTC is enhanced. In contrast, large-scale cavity-QED experiments with BECs operating with $N \sim 10^4-10^5$ will likely benefit from the rigidity of the thermodynamic limit.  Independent of the number of atoms, the observation of the non-trivial effects discussed here and in Ref.~\cite{Gong2018} requires the existence of a mean-field superradiant transition, which does not exist in most cases due to the inclusion of the $A^{2}$ term.  However non-photonic platforms where this term is absent~\cite{Kono}, as well as systems such as  transversally pumped Bose-Einstein condensates coupled to an optical cavity~\cite{Baumann} where this term is replaced by a shift term that is small in the vicinity of the transition, might provide avenues to observe the effects predicted in this work.

In conclusion, we have mapped the finite-size stability of DTC order in the open Dicke model, revealing a non-trivial non-monotonic approach to the $N = \infty$ behavior, as well as regimes of very long lived DTC stability. We have demonstrated that a stochastic version of the TWA provides a quantitative means to describe the dynamics of such non-mean field cases for long times, a fact that should motivate widespread application of this approach in both standard and driven open cavity systems.  Future work will be devoted to assessing the fate of DTC order on even longer time scales in the pre-thermal regime, as well as investigating related models which should display the effects exhibited here in realistic experimental settings.

\begin{acknowledgments}
This work was performed with support from the U.S.\ Department of Energy, Office of Science, Office of Advanced Scientific Computing Research, Scientific Discovery through Advanced Computing (SciDAC) program, under Award No. DE-SC0022088. We thank Jonathan Keeling, Andrew Millis, Anqi Mu, and Nathan Ng for insightful discussions.  In particular, we thank Nathan Ng for pointing us to Ref.~\cite{Hosseinabadi2025}. The code and data supporting this publication can be accessed at \url{https://github.com/nguye66h/qed-dtc}.
\end{acknowledgments}

\bibliography{ref}

@article{Gong2018,
  title = {Discrete Time-Crystalline Order in Cavity and Circuit QED Systems},
  volume = {120},
  ISSN = {1079-7114},
  url = {http://dx.doi.org/10.1103/PhysRevLett.120.040404},
  DOI = {10.1103/physrevlett.120.040404},
  number = {4},
  journal = {Physical Review Letters},
  publisher = {American Physical Society (APS)},
  author = {Gong,  Zongping and Hamazaki,  Ryusuke and Ueda,  Masahito},
  year = {2018},
  month = jan 
}

@article{Rzazewski,
  title = {Phase Transitions, Two-Level Atoms, and the ${A}^{2}$ Term},
  volume = {35},
  ISSN = {},
  url = {https://doi.org/10.1103/PhysRevLett.35.432},
  DOI = {10.1103/PhysRevLett.35.432},
  number = {35},
  journal = {Physical Review Letters},
  publisher = {American Physical Society (APS)},
  author = {Rzażewski, K and Wódkiewicz, K and Żakowicz, W},
  year = {1975},
  month = oct, 
  pages = {432-434}
}

@article{yao,
  title = {Discrete Time Crystals},
  volume = {11},
  ISSN = {},
  url = {https://doi.org/10.1146/annurev-conmatphys-031119-050658},
  DOI = {},
  number = {1},
  journal = {Annual Review of Condensed Matter Physics},
  publisher = {Annual Reviews},
  author = {Else, Dominic V. and Monroe, Christopher and Nayak, Chetan and Yao, Norman Y.},
  year = {2020},
  month = oct, 
  pages = {467-499}
}

@article{Hepp,
  title = {On the Superradiant Phase Transition for Molecules in a Quantized Radiation Field: Dicke Maser Model},
  volume = {76},
  ISSN = {},
  url = {},
  DOI = {10.1016/0003-4916(73)90039-0},
  number = {},
  journal = {Annals of Physics},
  publisher = {},
  author = {Hepp, K and Lieb, E H},
  year = {1973},
  month = oct, 
  pages = {360–404}
}

@article{Kirton2017,
  title = {Suppressing and Restoring the Dicke Superradiance Transition by Dephasing and Decay},
  volume = {118},
  ISSN = {1079-7114},
  url = {http://dx.doi.org/10.1103/PhysRevLett.118.123602},
  DOI = {10.1103/physrevlett.118.123602},
  number = {12},
  journal = {Physical Review Letters},
  publisher = {American Physical Society (APS)},
  author = {Kirton,  Peter and Keeling,  Jonathan},
  year = {2017},
  month = mar 
}

@article{Johansson2013,
  title = {QuTiP 2: A Python framework for the dynamics of open quantum systems},
  volume = {184},
  ISSN = {0010-4655},
  url = {http://dx.doi.org/10.1016/j.cpc.2012.11.019},
  DOI = {10.1016/j.cpc.2012.11.019},
  number = {4},
  journal = {Computer Physics Communications},
  publisher = {Elsevier BV},
  author = {Johansson,  J.R. and Nation,  P.D. and Nori,  Franco},
  year = {2013},
  month = apr,
  pages = {1234–1240}
}

@article{Johansson2012,
  title = {QuTiP: An open-source Python framework for the dynamics of open quantum systems},
  volume = {183},
  ISSN = {0010-4655},
  url = {http://dx.doi.org/10.1016/j.cpc.2012.02.021},
  DOI = {10.1016/j.cpc.2012.02.021},
  number = {8},
  journal = {Computer Physics Communications},
  publisher = {Elsevier BV},
  author = {Johansson,  J.R. and Nation,  P.D. and Nori,  Franco},
  year = {2012},
  month = aug,
  pages = {1760–1772}
}

@article{Dalibard1992,
  title = {Wave-function approach to dissipative processes in quantum optics},
  volume = {68},
  ISSN = {0031-9007},
  url = {http://dx.doi.org/10.1103/PhysRevLett.68.580},
  DOI = {10.1103/physrevlett.68.580},
  number = {5},
  journal = {Physical Review Letters},
  publisher = {American Physical Society (APS)},
  author = {Dalibard,  Jean and Castin,  Yvan and Mølmer,  Klaus},
  year = {1992},
  month = feb,
  pages = {580–583}
}

@article{Zhang2017,
  title = {Observation of a discrete time crystal},
  volume = {543},
  ISSN = {1476-4687},
  url = {http://dx.doi.org/10.1038/nature21413},
  DOI = {10.1038/nature21413},
  number = {7644},
  journal = {Nature},
  publisher = {Springer Science and Business Media LLC},
  author = {Zhang,  J. and Hess,  P. W. and Kyprianidis,  A. and Becker,  P. and Lee,  A. and Smith,  J. and Pagano,  G. and Potirniche,  I.-D. and Potter,  A. C. and Vishwanath,  A. and Yao,  N. Y. and Monroe,  C.},
  year = {2017},
  month = mar,
  pages = {217–220}
}

@article{Baumann,
  title = {Dicke Quantum Phase Transition With a Superfluid Gas in an Optical Cavity},
  volume = {464},
  ISSN = {1476-4687},
  url = {},
  DOI = {},
  number = {7644},
  journal = {Nature},
  publisher = {Springer Science and Business Media LLC},
  author = {Baumann, K and Guerlin, C and Brennecke, F and Esslinger, T},
  year = {2010},
  month = april,
  pages = {1301–1306}
}

@article{Choi2017,
  title = {Observation of discrete time-crystalline order in a disordered dipolar many-body system},
  volume = {543},
  ISSN = {1476-4687},
  url = {http://dx.doi.org/10.1038/nature21426},
  DOI = {10.1038/nature21426},
  number = {7644},
  journal = {Nature},
  publisher = {Springer Science and Business Media LLC},
  author = {Choi,  Soonwon and Choi,  Joonhee and Landig,  Renate and Kucsko,  Georg and Zhou,  Hengyun and Isoya,  Junichi and Jelezko,  Fedor and Onoda,  Shinobu and Sumiya,  Hitoshi and Khemani,  Vedika and von Keyserlingk,  Curt and Yao,  Norman Y. and Demler,  Eugene and Lukin,  Mikhail D.},
  year = {2017},
  month = mar,
  pages = {221–225}
}

@article{Smits2018,
  title = {Observation of a Space-Time Crystal in a Superfluid Quantum Gas},
  volume = {121},
  ISSN = {1079-7114},
  url = {http://dx.doi.org/10.1103/PhysRevLett.121.185301},
  DOI = {10.1103/physrevlett.121.185301},
  number = {18},
  journal = {Physical Review Letters},
  publisher = {American Physical Society (APS)},
  author = {Smits,  J. and Liao,  L. and Stoof,  H. T. C. and van der Straten,  P.},
  year = {2018},
  month = oct 
}

@article{Keler2021,
  title = {Observation of a Dissipative Time Crystal},
  volume = {127},
  ISSN = {1079-7114},
  url = {http://dx.doi.org/10.1103/PhysRevLett.127.043602},
  DOI = {10.1103/physrevlett.127.043602},
  number = {4},
  journal = {Physical Review Letters},
  publisher = {American Physical Society (APS)},
  author = {Keßler,  Hans and Kongkhambut,  Phatthamon and Georges,  Christoph and Mathey,  Ludwig and Cosme,  Jayson G. and Hemmerich,  Andreas},
  year = {2021},
  month = jul 
}

@article{Taheri2022,
  title = {All-optical dissipative discrete time crystals},
  volume = {13},
  ISSN = {2041-1723},
  url = {http://dx.doi.org/10.1038/s41467-022-28462-x},
  DOI = {10.1038/s41467-022-28462-x},
  number = {1},
  journal = {Nature Communications},
  publisher = {Springer Science and Business Media LLC},
  author = {Taheri,  Hossein and Matsko,  Andrey B. and Maleki,  Lute and Sacha,  Krzysztof},
  year = {2022},
  month = feb 
}

@article{Kono,
  title = {Observation of the Magnonic Dicke Superradiant Phase Transition},
  volume = {11},
  ISSN = {},
  url = {},
  DOI = {DOI: 10.1126/sciadv.adt1691},
  number = {},
  journal = {Science Advances},
  publisher = {American Association for the Advancement of Science (AAAS)},
  author = {Kim, D and Dasgupta, S and Ma, X and Park, J -M and Wei, H -T and Luo, L and Doumani, J and Yang, W and Cheng, D and Kim, R H J and Everitt, H O and Kimura, S and Nojiri, H and Wang, J and Cao, S and Bamba, M and Hazzard, K R A and Kono, J},
  year = {2025},
  month = aug,
  pages = {eadt1691}
}

@misc{Kerber,
    author = "Kerber, J and Ritsch, H and Ostermann, L",
    title = "{The cumulants expansion approach: The good, the bad and the ugly}",
    eprint = "2511.20115",
    archivePrefix = "arXiv",
    primaryClass = "quant-ph",
    month = "11",
    year = "2025"
}

@article{Hosseinabadi2025,
  title = {User-Friendly Truncated Wigner Approximation for Dissipative Spin Dynamics},
  volume = {6},
  ISSN = {2691-3399},
  url = {http://dx.doi.org/10.1103/1wwv-k7hg},
  DOI = {10.1103/1wwv-k7hg},
  number = {3},
  journal = {PRX Quantum},
  publisher = {American Physical Society (APS)},
  author = {Hosseinabadi,  Hossein and Chelpanova,  Oksana and Marino,  Jamir},
  year = {2025},
  month = Sept 
}

@article{Kirton2018,
  title = {Introduction to the Dicke Model: From Equilibrium to Nonequilibrium,  and Vice Versa},
  volume = {2},
  ISSN = {2511-9044},
  url = {http://dx.doi.org/10.1002/qute.201800043},
  DOI = {10.1002/qute.201800043},
  number = {1–2},
  journal = {Advanced Quantum Technologies},
  publisher = {Wiley},
  author = {Kirton,  Peter and Roses,  Mor M. and Keeling,  Jonathan and Dalla Torre,  Emanuele G.},
  year = {2018},
  month = oct 
}

\clearpage
\newpage
\setcounter{page}{1}
\setcounter{equation}{0}
\setcounter{section}{0}
\onecolumngrid

\appendix
\renewcommand{\thefigure}{(SM \arabic{figure})}
\setcounter{figure}{0}
\renewcommand{\theequation}{S\arabic{equation}}
\setcounter{equation}{0}

\begin{center}
\Large Supplementary Materials for ``Fate of Discrete Time Crystals in Finite-Size Cavity QED Systems''
\end{center}

\vspace*{2em}

This Supplement
\begin{itemize}
    \item shows the model and notation used in the paper;
    \item derives the full and partial second-order cumulant expansions (FSC, PSC) and shows why they are not reliable for this problem. This motivates the use of the truncated Wigner approximation (TWA) in the main text;
    \item gives the TWA equations of motion and initial-state sampling used for all TWA results, and shows that the quantum noise from cavity loss scales as $1/N$;
    \item derives the linearized Floquet map whose multipliers are plotted in Fig.~4(a) in the main text;
    \item explains the mechanism behind the non-monotonic dependence of the DTC stability on $N$ reported in the main text.
\end{itemize}

\newpage

\section{MODEL \& NOTATION} \label{sec:model}
\subsection{Model}
We consider $N$ two-level atoms (Pauli operators $\hat\sigma^{(i)}_{x,y,z}$, $i=1,\dots,N$) coupled to a single lossy cavity mode $\hat{a}$ ($\hbar=1$).
\begin{align}
  H(t) &= \omega_c \hat{a}^\dagger \hat{a} + \frac{\omega}{2} \sum_{i=1}^N \sigma^{(i)}_z + \frac{\lambda(t)}{\sqrt{N}} (\hat{a}^\dagger + \hat{a}) \sum_{i=1}^N \sigma^{(i)}_x, \label{eq:H}\\
  \dot{\hat\rho} &= -i[\hat H(t),\hat\rho] + \kappa\left(\hat{a}\hat\rho\hat{a}^\dagger - \tfrac12\{\hat{a}^\dagger\hat{a},\hat\rho\}\right). \label{eq:lindblad}
\end{align}
Here $\omega_c$ is the cavity frequency, $\omega$ the atomic transition frequency, and $\kappa$ the decay rate of the cavity. The coupling is modulated as a square wave of period $T$ and amplitude $\lambda_0$ ($\lambda(t) = \lambda_0$ for $0 \le t~(\bmod~T) < T/2$ and $\lambda(t) = 0$ for $T/2 \le t~(\bmod~T) < T$ with $T=2\pi$). The detuning parameter $\epsilon$ is defined as $\epsilon=(\omega-\omega_c)/2$.
\subsection{$\mathbb{Z}_2$ symmetry and the DTC}
The parity operator
\begin{align}
    \hat{\Pi} = \exp\! \Big[i\pi \Big( \hat{a}^\dagger \hat{a} + \sum_i \tfrac{1}{2} (\sigma_z^{(i)}+1) \Big) \Big] \label{eq:parity_op}
\end{align}
maps $\hat{a}\to -\hat{a}$, $\sigma^{(i)}_{x,y} \to -\sigma^{(i)}_{x,y}$ and $\sigma^{(i)}_z\to \sigma^{(i)}_z$.
It commutes with $H(t)$ at all times and with the dissipator, so Eqs.~\eqref{eq:H}--\eqref{eq:lindblad} are $\mathbb{Z}_2$ symmetric. 
In the superradiant regime, the mean-field dynamics has two symmetry-broken states related by $\hat{\Pi}$, with opposite signs of $\langle \hat{a} \rangle$ and $\langle \sigma_{x,y}^{(i)} \rangle$. The DTC is the response in which the drive switches the system between these two states every period, so that $\langle \sigma^{(i)}_x \rangle$ changes sign at each stroboscopic time $t=nT$.
\section{FULL AND PARTIAL SECOND-ORDER CUMULANT EXPANSIONS} \label{sec:cumulant}
A common inexpensive method for large-$N$ cavity QED is the second-order cumulant expansion. In this section, we derive the full (FSC) and partial (PSC) versions for Eqs.\eqref{eq:H}--\eqref{eq:lindblad}. We then benchmark both against exact results and show that they can be numerically unstable and give initial-state-dependent steady states.
\subsection{Method}
The Heisenberg equations generated by Eqs.~\eqref{eq:H}--\eqref{eq:lindblad} couple moments of order $n$ to moments of order $n+1$. We close the hierarchy at second order by setting all third-order cumulants to zero.
For any three operators $A, B, C$, this means that,
\begin{align}
    \langle ABC \rangle \approx \langle AB \rangle \langle C \rangle + \langle AC \rangle \langle B \rangle + \langle BC \rangle \langle A \rangle - 2 \langle A \rangle \langle B \rangle \langle C \rangle .
\end{align}
The initial states and dynamics we consider are permutation symmetric.
Therefore, all single-site quantities are independent of the site index, and all two-site correlators are independent of the pair ($i \neq j$). 
We write
\begin{align}
    \langle \sigma_\mu \rangle \equiv \langle \sigma_\mu^{(i)} \rangle, \qquad
    \langle \hat{a} \sigma_\mu \rangle \equiv \langle \hat{a} \sigma_\mu^{(i)} \rangle, \qquad
    C_{\mu\nu} \equiv \langle \sigma_\mu^{(i)} \sigma_\nu^{(j)} \rangle.
\end{align}
\subsection{Full second-order cumulant expansion (FSC)}
The FSC equations are
\begin{align}
    \frac{d\langle\hat{a}\rangle}{dt}&=-\left( i\omega_c+\frac{\kappa}{2} \right) \langle{\hat{a}}\rangle -i\lambda(t)\sqrt{N}  \langle{\sigma_x}\rangle,\\
    \frac{d\langle{\sigma_x}\rangle}{dt}&=- \omega \langle{\sigma_y}\rangle,\\
    \frac{d\langle{\sigma_y}\rangle}{dt}&= \omega \langle{\sigma_x}\rangle-4\frac{\lambda(t)}{\sqrt{N}}\Re\left[\langle{\hat{a}\sigma_z}\rangle\right],\\
    \frac{d\langle{\sigma_z}\rangle}{dt}&= 4\frac{\lambda(t)}{\sqrt{N}}\Re\left[\langle{\hat{a}\sigma_y}\rangle\right],\\
    \frac{d\langle{\hat{a}\sigma_x}\rangle}{dt}&=-\left( i\omega_c+\frac{\kappa}{2} \right) \langle{\hat{a}\sigma_x}\rangle-i\frac{\lambda(t)}{\sqrt{N}}(N-1)C_{xx}-i\frac{\lambda(t)}{\sqrt{N}}-\omega \langle{\hat{a}\sigma_y}\rangle,\\
    \frac{d\langle{\hat{a}\sigma_y}\rangle}{dt}&=-\left( i\omega_c+\frac{\kappa}{2} \right) \langle{\hat{a}\sigma_y}\rangle-i\frac{\lambda(t)}{\sqrt{N}}(N-1)C_{xy}-\frac{\lambda(t)}{\sqrt{N}}\langle{\sigma_z}\rangle+\omega\langle{a\sigma_x}\rangle \nonumber\\
    &-2\frac{\lambda(t)}{\sqrt{N}}\left( 2\langle\hat{a}\rangle\langle\hat{a}\sigma_z\rangle+\langle\sigma_z\rangle\langle\hat{a}\hat{a}\rangle-2\langle\hat{a}\rangle^2\langle\sigma_z\rangle 
    +  \langle \hat{a} \rangle \langle \hat{a} \sigma_z \rangle^* + \langle \hat{a} \rangle^* \langle \hat{a} \sigma_z \rangle  + \langle \sigma_z \rangle \langle \hat{a}^\dagger \hat{a}\rangle -2\langle \hat{a} \rangle \langle \hat{a} \rangle^* \langle \sigma_z \rangle \right),\\
    \frac{d\langle{\hat{a}\sigma_z}\rangle}{dt}&=-\left( i\omega_c+\frac{\kappa}{2} \right) \langle{\hat{a}\sigma_z}\rangle-i\frac{\lambda(t)}{\sqrt{N}}(N-1)C_{xz}+\frac{\lambda(t)}{\sqrt{N}}\langle{\sigma_y}\rangle \nonumber\\
    &+2\frac{\lambda(t)}{\sqrt{N}}\left( 2\langle\hat{a}\rangle\langle\hat{a}\sigma_y\rangle+\langle\sigma_y\rangle\langle\hat{a}\hat{a}\rangle-2\langle\hat{a}\rangle^2\langle\sigma_y\rangle
    +  \langle \hat{a} \rangle \langle \hat{a} \sigma_y \rangle^* + \langle \hat{a} \rangle^* \langle \hat{a} \sigma_y \rangle  + \langle \sigma_y \rangle \langle \hat{a}^\dagger \hat{a}\rangle -2\langle \hat{a} \rangle \langle \hat{a} \rangle^* \langle \sigma_y \rangle \right),\\
    \frac{dC_{xx}}{dt}&=-2\omega  C_{xy} ,\\
    \frac{dC_{yy}}{dt}&=2\omega C_{xy}-8\frac{\lambda(t)}{\sqrt{N}}\Re\left[\langle \sigma_y \rangle \langle \hat{a}\sigma_z \rangle+\langle \hat{a} \rangle C_{yz} + \langle \sigma_z \rangle \langle \hat{a}\sigma_y \rangle-2\langle \hat{a} \rangle \langle \sigma_y \rangle \langle \sigma_z \rangle\right],\\
    \frac{dC_{zz}}{dt}&=8\frac{\lambda(t)}{\sqrt{N}}\Re\left[\langle \sigma_z \rangle \langle \hat{a}\sigma_y \rangle+\langle \hat{a} \rangle C_{yz} + \langle \sigma_y \rangle \langle \hat{a}\sigma_z \rangle-2\langle \hat{a} \rangle \langle \sigma_y \rangle \langle \sigma_z \rangle\right],\\
    \frac{dC_{xy}}{dt}&=\omega \left(C_{xx}- C_{yy}\right)-4\frac{\lambda(t)}{\sqrt{N}}\Re\left[ \langle \sigma_x \rangle \langle \hat{a}\sigma_z \rangle+\langle \hat{a} \rangle C_{xz} + \langle \sigma_z \rangle \langle \hat{a}\sigma_x \rangle-2\langle \hat{a} \rangle \langle \sigma_x \rangle \langle \sigma_z \rangle \right],\\
    \frac{dC_{xz}}{dt}&=-\omega C_{yz}+4\frac{\lambda(t)}{\sqrt{N}}\Re\left[ \langle \sigma_x \rangle \langle \hat{a}\sigma_y \rangle+\langle \hat{a} \rangle C_{xy} + \langle \sigma_y \rangle \langle \hat{a}\sigma_x \rangle-2\langle \hat{a} \rangle \langle \sigma_x \rangle \langle \sigma_y \rangle \right],\\
    \frac{d C_{yz}}{dt}&= \omega C_{xz}+4\frac{\lambda(t)}{\sqrt{N}}\Re \left[ \langle \hat{a} \rangle \left(C_{yy} - C_{zz}\right)+2\left(\langle \sigma_y \rangle \langle \hat{a} \sigma_y \rangle -\langle \sigma_z \rangle \langle \hat{a} \sigma_z \rangle \right)
    -2\langle \hat{a} \rangle \left(\langle \sigma_y \rangle ^2 - \langle \sigma_z \rangle ^2\right) \right],\\
    \frac{d\langle{\hat{a}^\dagger \hat{a}}\rangle}{dt}&=-\kappa \langle \hat{a}^\dagger \hat{a} \rangle-2\lambda(t)\sqrt{N}\Im \left[ \langle \hat{a} \sigma_x \rangle \right],\\
    \frac{d \langle{\hat{a}\hat{a}}\rangle}{dt}&=-2\left(i\omega_c+\frac{\kappa}{2} \right) \langle \hat{a} \hat{a} \rangle-2i\lambda(t)\sqrt{N} \langle \hat{a} \sigma_x \rangle.
\end{align}
\subsection{Partial second-order cumulant expansion (PSC)}
The parity operator $\hat{\Pi}$ (Eq.~\eqref{eq:parity_op}) commutes with $H(t)$ and the Lindblad dissipator. Therefore, in a thermodynamically equilibrium state, every moment that is odd under $\hat{\Pi}$ vanishes:
\begin{align}
    \langle \hat{a} \rangle = \langle \sigma_x \rangle=\langle \sigma_y \rangle=\langle \hat{a} \sigma_z \rangle=C_{xz} = C_{yz}=0.
\end{align}
In addition, if the initial state is set to be a vacuum cavity with all spins in the $+1$ eigenstate of $\hat\sigma_z$, i.e.\ $\langle \hat{a} \rangle=\langle\sigma_x\rangle=\langle\sigma_y\rangle=0$ and $\langle\sigma_z\rangle=1$, the dynamics stays in this parity-symmetric subspace since $\hat{\Pi}$ is conserved. Setting the $\mathbb{Z}_2$-symmetry-breaking variables to zero, we arrive at a simpler set of differential equations as follows:
\begin{align}
    \frac{d\langle{\sigma_z}\rangle}{dt}&= 4\frac{\lambda(t)}{\sqrt{N}}\Re\left[\langle{\hat{a}\sigma_y}\rangle\right],\\
\frac{d\langle{\hat{a}\sigma_x}\rangle}{dt}&=-\left( i\omega_c+\frac{\kappa}{2} \right) \langle{\hat{a}\sigma_x}\rangle-i\frac{\lambda(t)}{\sqrt{N}}\Big[(N-1)   C_{xx}+1\Big]-\omega \langle{\hat{a}\sigma_y}\rangle,\\
\frac{d\langle{\hat{a}\sigma_y}\rangle}{dt}&=-\left( i\omega_c+\frac{\kappa}{2} \right) \langle{\hat{a}\sigma_y}\rangle-i\frac{\lambda(t)}{\sqrt{N}}(N-1)C_{xy}-\frac{\lambda(t)}{\sqrt{N}}\langle{\sigma_z}\rangle\left(1+2\langle\hat{a}\hat{a}\rangle+2\langle\hat{a}^\dagger\hat{a}\rangle\right)+\omega \langle{\hat{a}\sigma_x}\rangle,\\
    \frac{d C_{xx}}{dt}&=-2\omega C_{xy},\\
    \frac{dC_{yy}}{dt}&=2\omega  C_{xy}-8\frac{\lambda(t)}{\sqrt{N}}\langle \sigma_z \rangle\Re\left[ \langle \hat{a}\sigma_y \rangle\right],\\
    \frac{d C_{zz}}{dt}&=8\frac{\lambda(t)}{\sqrt{N}}\langle \sigma_z \rangle\Re\left[ \langle \hat{a}\sigma_y \rangle\right],\\
    \frac{dC_{xy}}{dt}&=\omega \left(C_{xx}- C_{yy}\right)-4\frac{\lambda(t)}{\sqrt{N}}\langle \sigma_z \rangle\Re\left[  \langle \hat{a}\sigma_x \rangle \right],\\
    \frac{d\langle{\hat{a}^\dagger \hat{a}}\rangle}{dt}&=-\kappa \langle \hat{a}^\dagger \hat{a} \rangle-2\lambda(t)\sqrt{N}\Im \left[ \langle \hat{a} \sigma_x \rangle \right],\\
    \frac{d \langle{\hat{a}\hat{a}}\rangle}{dt}&=-2\left(i\omega_c+\frac{\kappa}{2} \right) \langle \hat{a} \hat{a} \rangle-2i\lambda(t)\sqrt{N} \langle \hat{a} \sigma_x \rangle.
\end{align}
For parity-symmetric initial states, FSC and PSC are therefore identical. The PSC cannot describe symmetry-broken states, such as the DTC, which by definition has $\langle \sigma_x \rangle \neq 0$.
\subsection{Benchmark against exact dynamics}
\begin{figure}
    \centering
    \includegraphics[width=0.5\linewidth]{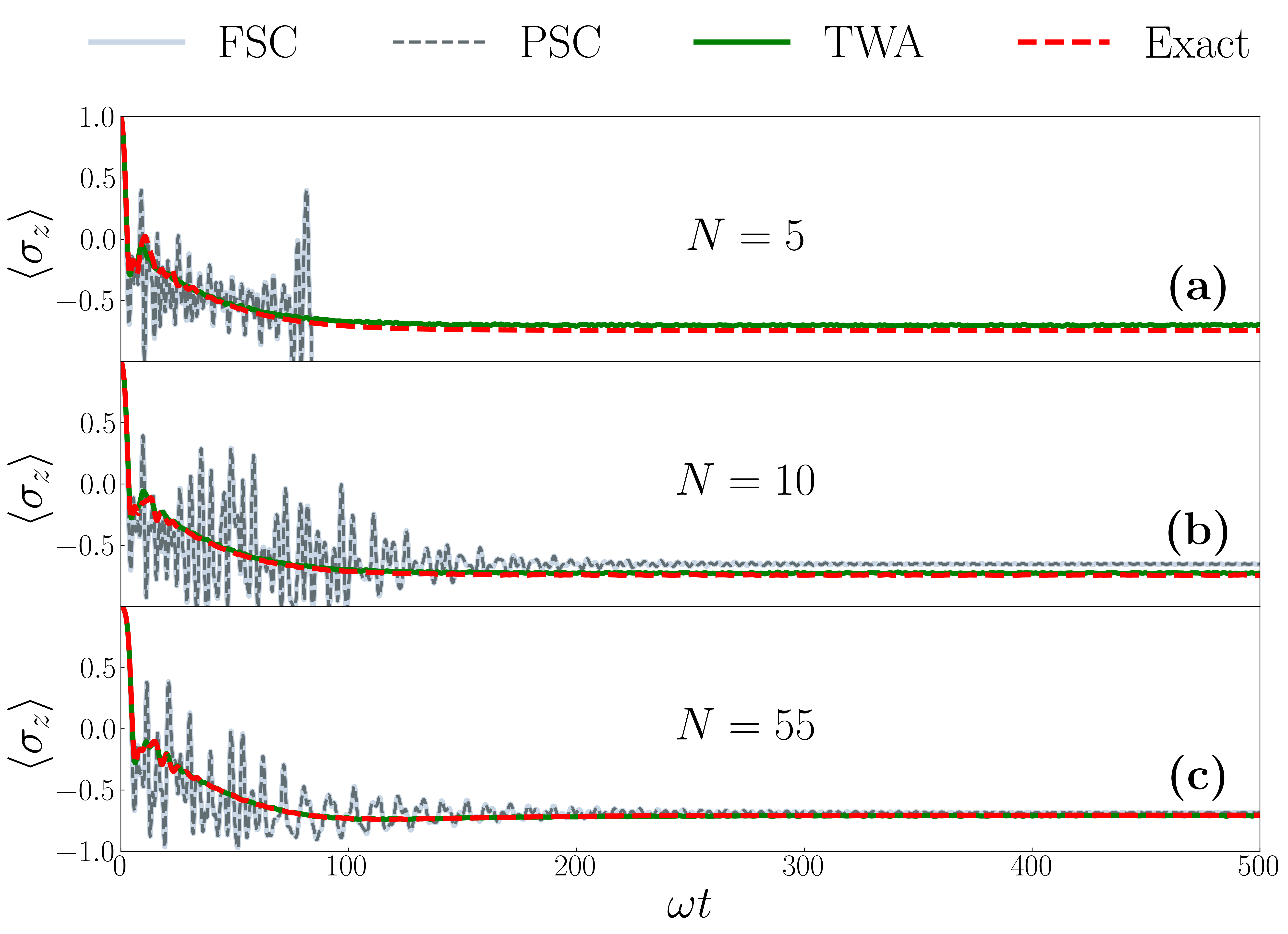}
    \caption{Time evolution of $\langle \sigma_z(t) \rangle$ from FSC (light gray solid), PSC (gray dashed), TWA (green), and exact (red) calculations for (a) $N=5$, (b) $N=10$, and (c) $N=55$. Initial state: vacuum cavity, all spins in the $+1$ eigenstate of $\sigma_z$. Parameters: $\epsilon=0.0, \kappa=0.05$, and $\lambda_0=0.6$ without coupling modulation.}
    \label{fig:TWAvsCEvsExact_sz}
\end{figure}
Fig.~\ref{fig:TWAvsCEvsExact_sz} compares the two cumulant expansions with the TWA and exact numerics for the parity-symmetric initial state. The PSC has been reported to give accurate steady-state values~\cite{Kirton2017}.
However, it shows large erratic oscillations before relaxing, and in particular at small $N=5$, fails to converge to a true steady state.
This is an instability of the closed equations themselves, not of the integrator.
By contrast, the TWA reproduces the exact dynamics closely at all three system sizes.
\begin{figure}
    \centering
    \includegraphics[width=0.5\linewidth]{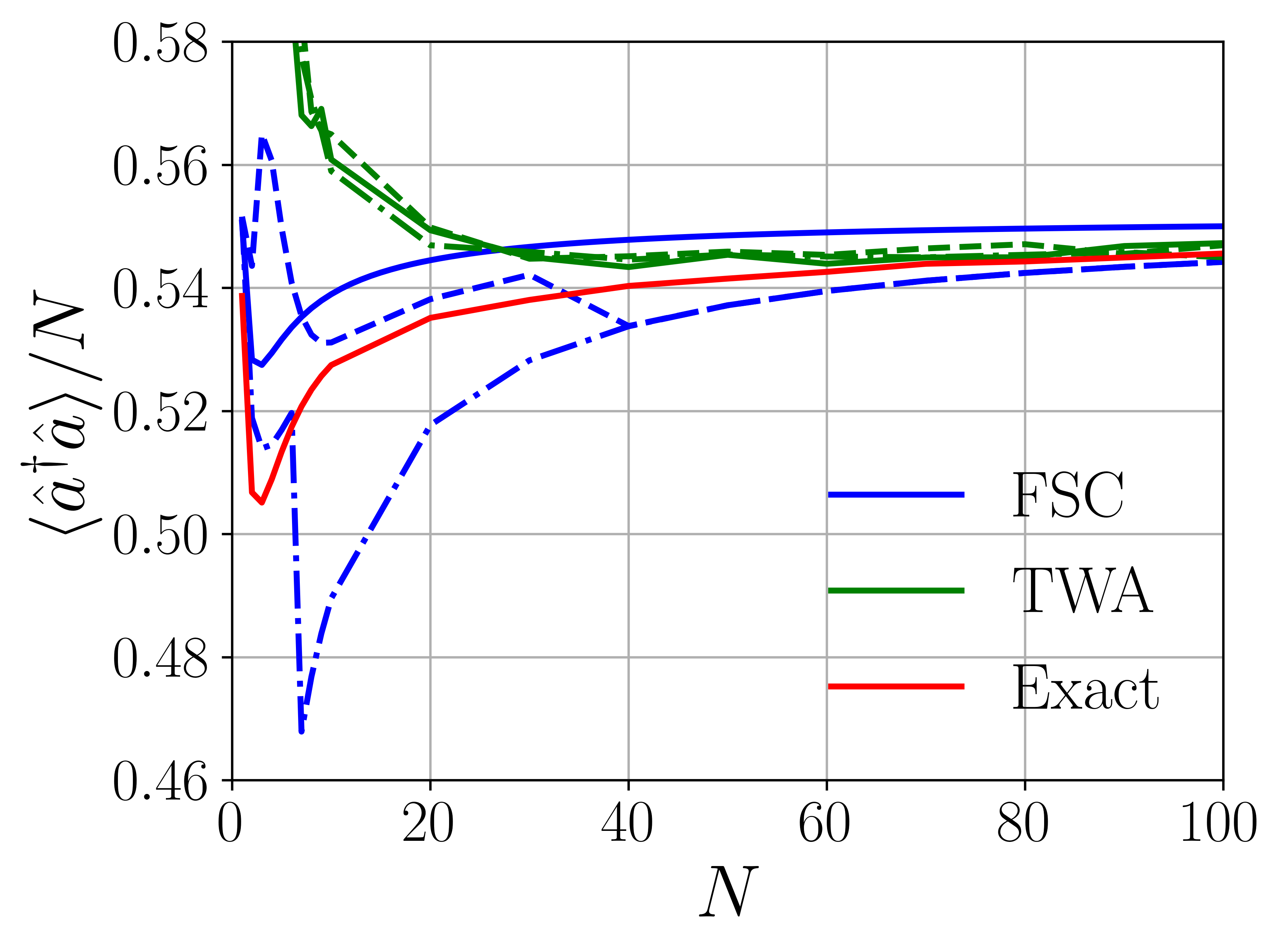}
    \caption{Steady-state cavity photon number per atom, $\langle \hat{a}^\dagger \hat{a} \rangle/N$, versus $N$ from exact (red), FSC (blue), and TWA (green) calculations.
    Line styles denote the initial spin state (with a vacuum cavity): dashed, dash-dotted, and solid correspond to the $+1$ eigenstates of $\sigma_x,~\sigma_y$, and $\sigma_z$, respectively.
    Parameters: $\epsilon=0.0, \kappa=1.0$, $\lambda_0=0.9$. This parameter regime is selected since the PSC has been reported to perform well here without numerical instability~\cite{Kirton2017}. }
    \label{fig:number_avg}
\end{figure}
Fig.~\ref{fig:number_avg} compares steady states for three different initial states. The exact steady state does not depend on the initial state. The FSC, however, relaxes to different values depending on where it starts, which is an artifact of the truncation. The TWA shows negligible dependence on the initial state and converges to exact results with increasing $N$.

\section{TRUNCATED WIGNER APPROXIMATION} \label{sec:twa}
This section derives TWA from the Lindblad equation and identifies which terms are kept, gives the stochastic equations and initial-state sampling used for every TWA result in the main text, and confirms numerically that the quantum noise from cavity loss contributes at order $1/N$.
\subsection{Phase-space representation}
Using the Weyl-Wigner correspondence, Eq.~\eqref{eq:lindblad} maps to a partial differential equation for the Wigner function $\rho_W(\alpha,\alpha^*,\vec{S})$:
\begin{equation}
    \dot{\rho}_W = [[H_W(t), \rho_W]] + \kappa \left( \alpha\star\rho_W \star\alpha^* - \frac{1}{2}\{\{\alpha^* \star\alpha, \rho_W\}\} \right), \label{eq:wigner_eq}
\end{equation}
with the Weyl symbol of Eq.~\eqref{eq:H},
\begin{align}
H_W(t) &= \omega_c |\alpha|^2 + \frac{\omega}{2} \sum_is^{(i)}_z + \frac{\lambda(t)}{\sqrt{N}} (\alpha^* + \alpha) \sum_i s^{(i)}_x.
\label{eq:HW}
\end{align}
For a canonical pair $(x,p)$, the Moyal star product, Moyal bracket, anti-commutator, and Poisson bracket are
\begin{align}
(A \star B)(x, p) &= A(x, p) \exp\left[ \frac{i}{2} \left( \overleftarrow{\partial_x}\overrightarrow{\partial_p} - \overleftarrow{\partial_p}\overrightarrow{\partial_x} \right) \right] B(x, p),\\
[[A, B]] &= -i (A \star B - B \star A) = 2 A(x, p) \sin\left[ \frac{1}{2} \left( \overleftarrow{\partial_x}\overrightarrow{\partial_p} - \overleftarrow{\partial_p}\overrightarrow{\partial_x} \right) \right] B(x, p),\\
\{\{A, B\}\} &= A \star B + B \star A=2 A(x, p) \cos\left( \frac{1}{2} \left( \overleftarrow{\partial_x}\overrightarrow{\partial_p} - \overleftarrow{\partial_p}\overrightarrow{\partial_x} \right) \right) B(x, p),\\
\{A,B\}_\text{P} &= \partial_x A \partial_p B - \partial_p A \partial_x B.
\end{align}
The cavity has the canonical pair $x=(\alpha+\alpha^*)/\sqrt{2},~p=(\alpha-\alpha^*)/(\sqrt{2}i)$. The spins are not canonical. Their classical variables obey the spin Poisson bracket $\{s^{(i)}_\mu,s^{(j)}_\nu\}=2\delta_{ij}\epsilon_{\mu\nu\gamma}s^{(i)}_\gamma$, and the corresponding phase-space expansion follows the same structure~\cite{Hosseinabadi2025}.

\subsection{TWA scaling with size $N$}
We first consider the unitary evolution that gives rise to the mean-field dynamics in TWA:
\begin{align}
    [[H_W(t),\rho_W]]=\underbrace{\{H_W(t),\rho_W\}_{\text{P}}}_{\text{Classical evolution}}+\underbrace{\mathcal{O}\left(\frac{1}{N^2}\right)}_{\text{Quantum coherence}}.
\end{align}
In contrast, the dissipative term gives rise to the deterministic drift and the stochastic noise terms in the TWA equations:
\begin{align}
    \kappa \left( \alpha\star\rho_W \star\alpha^* - \frac{1}{2}\{\{\alpha^* \star\alpha, \rho_W\}\} \right)\approx \underbrace{\frac{\kappa}2{}\left(\frac{\partial}{\partial\alpha}(\alpha\rho_W)+\frac{\partial}{\partial\alpha^*}(\alpha^*\rho_W)\right)}_{\text{Drift}}+\underbrace{\kappa\frac{\partial^2\rho_W}{\partial\alpha\partial\alpha^*}}_{\text{Diffusion}} \label{eqn:diss_term}
\end{align}
Given the fact that $\alpha\propto\sqrt{N}$, we have the following scaling comparisons:
\begin{itemize}
    \item Classical evolution and drift: $\mathcal{O}(1)$
    \item Diffusion: $\mathcal{O}(1/N)$
    \item Quantum dynamical corrections: $\mathcal{O}(1/N^2)$
\end{itemize}
The TWA keeps all terms up to and including the $O(1/N)$ diffusion and drops the $O(1/N^2)$ corrections.
\subsection{TWA equations of motion}
Each TWA trajectory obeys
\begin{align}
    \dot{\alpha} &= -i\omega_c \alpha - \frac{i\lambda(t)}{\sqrt{N}} \sum_i s_x^{(i)} - \frac{\kappa}{2}\alpha - \frac{\xi_\kappa}{2},\label{eqn:TWA0} \\
    \dot{s}_x^{(i)} &= -\omega s_y^{(i)},\label{eqn:TWA1} \\
    \dot{s}_y^{(i)} &= \omega s_x^{(i)} - \frac{4\lambda(t)}{\sqrt{N}} \Re[\alpha] s_z^{(i)}, \label{eqn:TWA2}\\
    \dot{s}_z^{(i)} &= \frac{4\lambda(t)}{\sqrt{N}} \Re[\alpha] s_y^{(i)}.\label{eqn:TWA3}
\end{align}
$\xi_\kappa$ is a complex Gaussian noise term centered at 0 satisfying $\overline{ \xi_\kappa(t) \xi_\kappa^*(t') } = 2\kappa \delta(t-t')$. 
The photon field $\alpha$ is sampled from the Gaussian Wigner function of the bosonic coherent state $|\alpha_0\rangle$, $W(\alpha)=\frac{2}{\pi}\exp{\left[-2\left|\alpha-\alpha_0\right|^2\right]}$, while each spin is sampled as $(s^{(i)}_x,s^{(i)}_y,s^{(i)}_z)=(\pm1,\pm1,+1)$ before being rotated to the desired initial state.

Because both the Hamiltonian and Lindblad dissipator yield dynamics that act directly on the collective spin, we can remove the index $i$ that distinguishes an individual atom in the ensemble and define the averaged spin variables as follows.
\begin{align}
    s_\mu = \frac{1}{N}\sum_{i=1}^N s_\mu^{(i)}.
\end{align}
\subsection{Numerical confirmation of the noise scaling}
\begin{figure}
    \centering
    \includegraphics[width=0.5\linewidth]{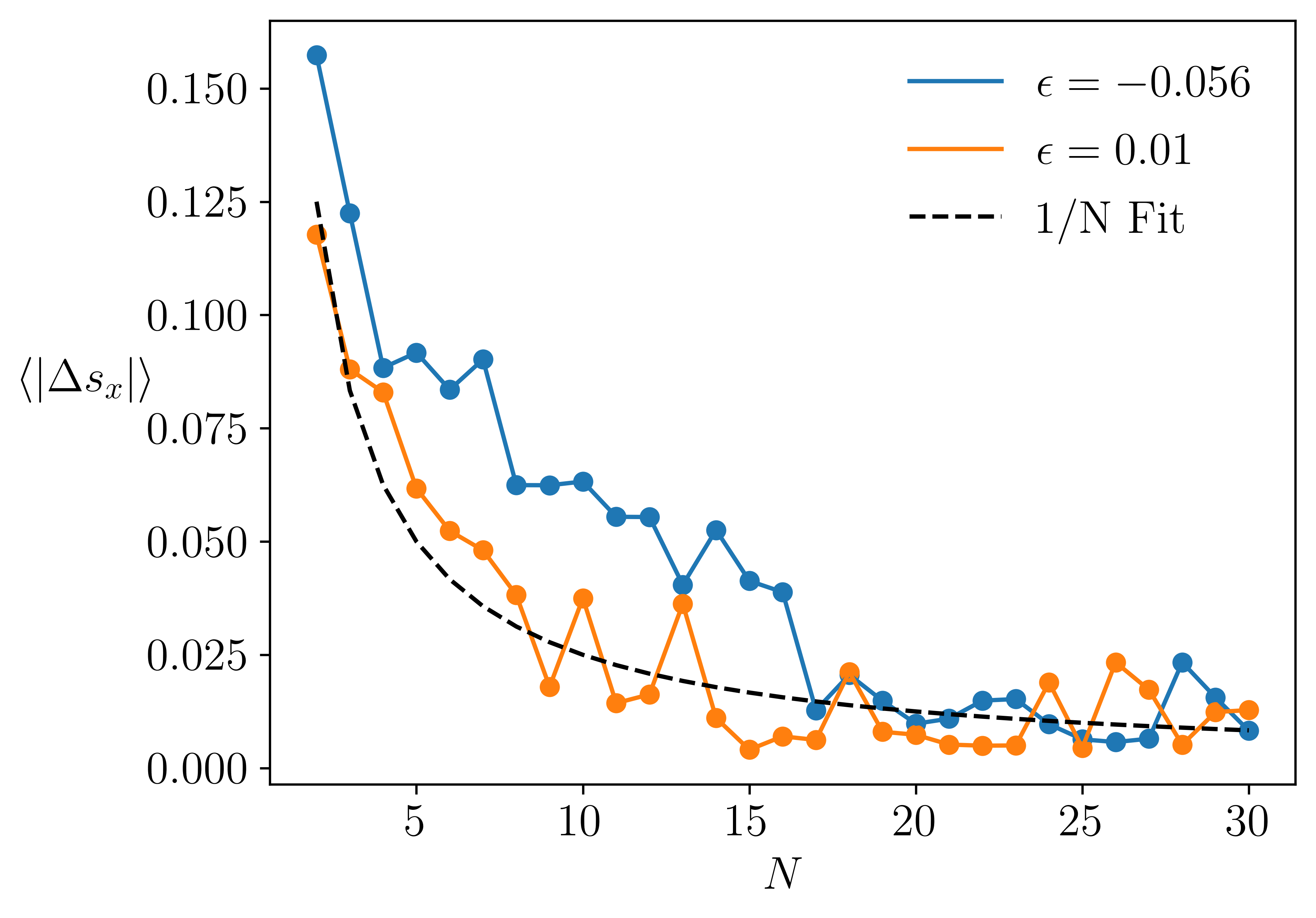}
    \caption{Effect of the cavity-loss noise in the TWA versus $N$. Plotted is the mean absolute difference between $\langle s_x \rangle$ computed with and without the noise term in Eq.~\eqref{eqn:TWA0}, averaged over the first 100 drive periods. Both runs use identical sampled initial conditions. Curves: $\epsilon=-0.056$ (blue), $\epsilon=0.01$ (orange), $1/N$ fit (black dashed).
Parameters: $\kappa=0.05,~\lambda_0=1.0$, and square-wave modulation of $\lambda(t)$.}
    \label{fig:noise_effect}
\end{figure}
Fig.~\ref{fig:noise_effect} confirms numerically that the noise contribution decreases as $1/N$ as expected from Eq.~\eqref{eqn:diss_term}. Note that removing the noise does not reduce the TWA to mean field. The sampled initial conditions still have a spread of $\propto 1/\sqrt{N}$, and at moderate $N$ this spread dominates the difference between the TWA and mean field.
\section{LINEARIZED FLOQUET ANALYSIS} \label{sec:linear}
Fig.~4(a) in the main text shows the magnitude of the largest Floquet multiplier as a function of $\epsilon$ and $\lambda_0$. This section derives the linear map from which those multipliers are computed and states its assumptions.
\subsection{Quasi-static longitudinal spin}
\begin{figure}
    \centering
    \includegraphics[width=\linewidth]{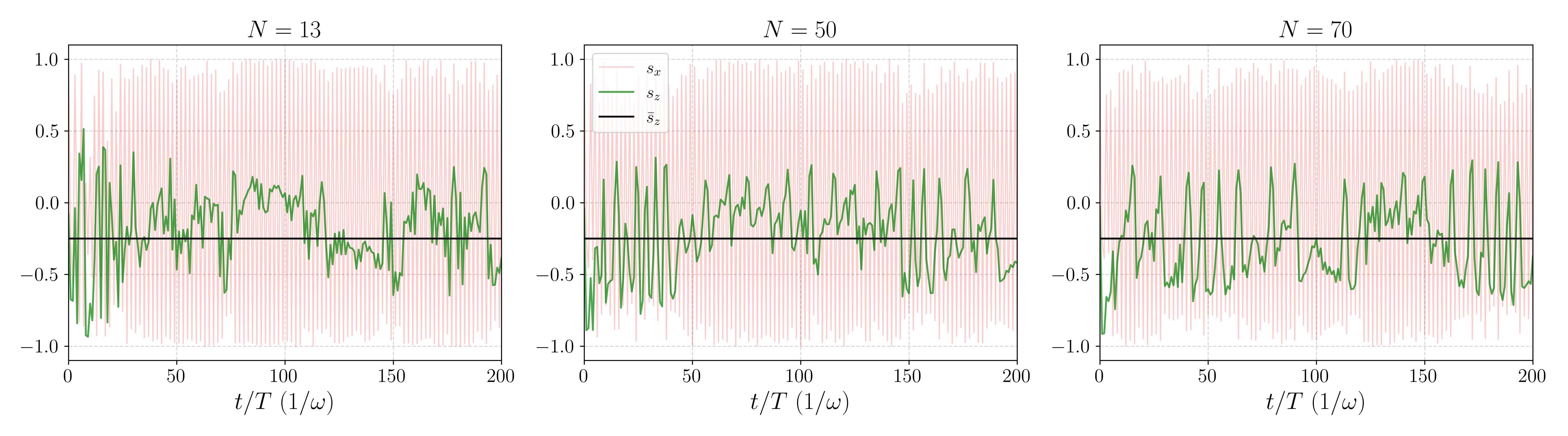}
    \caption{$s_x(nT)$ (red), $s_z(nT)$ (green), and the unmodulated steady-state value of $s_z$, $\bar s_z$ (black), over 200 drive periods for one TWA trajectory at (a) $N=13$, (b) $N=50$, and (c) $N=70$. Parameters: $\epsilon=0.03,~\kappa=0.05,~\lambda_0=1.0$}
    \label{fig:sz_timescale}
\end{figure}
In the regimes of interest, oscillations of the longitudinal spin component $s_z$ are smaller in both amplitude and frequency than those of the cavity field $\alpha$ and the transverse spin components $s_x$ and $s_y$ (Fig.~\ref{fig:sz_timescale}). 
To simplify our analysis, we substitute $s^{(i)}_z$ with $\bar{s}_z$, the mean-field steady state value of the unmodulated model with coupling $\lambda_0$. 
Setting the time derivatives in Eqs.~\eqref{eqn:TWA0}--\eqref{eqn:TWA3} to zero (without noise) gives, in the superradiant phase $\lambda_0 > \lambda_c$,
\begin{align}
    \bar{s}_z&=-\frac{\omega(\omega_c^2+\kappa^2/4)}{4\lambda_0^2\omega_c},\\
    \lambda_c&=\frac{1}{2}\sqrt{\frac{\omega}{\omega_c}\Big(\omega_c^2+\frac{\kappa^2}{4}\Big)}.
\end{align}
\subsection{Linearized Floquet map}
With $s_z^{(i)} \to \bar s_z$, Eqs.~\eqref{eqn:TWA0}--\eqref{eqn:TWA3} become linear. Averaging over trajectories removes the noise exactly because $\langle \xi_k \rangle =0$ and the equations are linear. 
The trajectory averages $a \equiv \langle \alpha \rangle/\sqrt{N}=u+iv,~\langle s_x \rangle$, and $\langle s_y \rangle$ then obey a closed, deterministic system $\dot{\bm x}=A(\lambda(t))\,\bm x$ with $\bm x = (u,v,\langle s_x \rangle, \langle s_y \rangle)^T$ and 
\begin{align}
  A(\lambda) =
  \begin{pmatrix}
    -\kappa/2 & \omega_c & 0 & 0\\
    -\omega_c & -\kappa/2 & -\lambda & 0\\
    0 & 0 & 0 & -\omega\\
    -4\lambda\bar s_z & 0 & \omega & 0
  \end{pmatrix}.
  \label{eq:A}
\end{align}
This simplification has two major implications:
\begin{itemize}
    \item Independence from initial samplings: Since the equations are now linear, the stability of the system is determined solely by the dynamical map itself, not by the specific initial conditions.
    \item Independence from $N$: In the rescaled variables, $A$ contains no $N$. The multipliers therefore give the mean-field ($N \to \infty)$ stability boundary.
    Finite-$N$ effects arise from nonlinearity with the initial state uncertainty and are treated in the next section.
\end{itemize}
For the coupling modulation, the monodromy matrix over one period is
\begin{align}
    M=e^{A(\lambda_0)T} e^{A(0)T}
\end{align}
\subsection{Floquet Multipliers}
The Floquet multipliers $\mu$ are the eigenvalues of $M$.
Since $M$ is real, its eigenvalues are real or come in complex-conjugate pairs.
We use the Floquet multiplier of the largest modulus as an order parameter:
\begin{itemize}
    \item $\mu \approx -1$ (real and negative, with $|\mu| \approx 1$): the state changes sign every period with almost no decay, i.e. robust DTC;
    \item $|\mu| < 1$: the period-doubled amplitude decays (decaying DTC);
    \item $|\mu| > 1$: the linearized dynamics is unstable.
\end{itemize}

\section{MECHANISM OF THE NON-MONOTONIC FINITE-$N$ SCALING} \label{sec:nonmono}
The main text reports that the DTC stability in both pre-thermal and transient regimes depend non-monotonically on $N$.
This section traces that behavior to two effects.
First, the finite-$N$ spread of TWA initial conditions straddles the basins of two competing stroboscopic attractors (Fig.~\ref{fig:mf_phase_space}). Second, in the transient regime, chaotic dynamics decorrelate the trajectories (Fig.~\ref{fig:lyapunov}). Here, the $\epsilon=0$ ($\epsilon=0.09$) regime is a representative of the pre-thermal (transient) dynamics. Both use $\kappa=0.05$ and $\lambda_0=1.0$.

\subsection{Two stroboscopic attractors and their basins}

\begin{figure}
    \centering
    \includegraphics[width=1.0\linewidth]{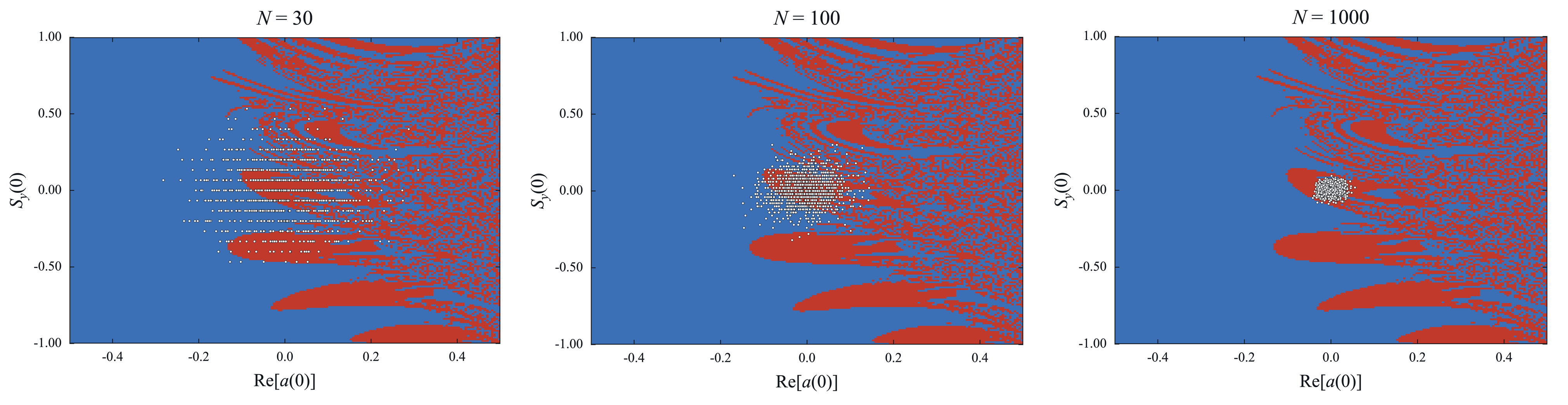}
    \caption{Basins of attraction of the two long-time stroboscopic attractors of the mean-field dynamics in the pre-thermal regime. 
    The two attractors are period-doubled orbits of opposite phases.
    Each pixel is a mean-field initial condition that varies only in $\Re{[a(0)]}$ (horizontal) and $S_y(0)$ (vertical); $S_x(0)$ is fixed at 1; all other variables are fixed at zero. The white dots represent TWA initial samples projected onto this plane for $N=30,100,$ and $1000$.  This figure only shows one slice in phase space with respect to initial parameters $\Re{[a(0)]}$ and $S_y(0)$; all other slice combinations show qualitatively similar features.
    Parameters: $\epsilon=0$, ~$\kappa=0.05$, ~$\lambda_0=1.0$.}
    \label{fig:mf_phase_space}
\end{figure}

For the parameters considered, the mean-field dynamics has two stroboscopic attractors, which are related by the $\mathbb{Z}_2$ symmetry and are period-doubled orbits of opposite phases. 
Their basins are separated by an intricate boundary.
Fig.~\ref{fig:mf_phase_space} shows this boundary in a single two-dimensional slice of the high-dimensional phase space.
The mean-field initial state lies in the red (in-phase) basin, but blue (out-of-phase) regions lie close by.
As observed, the ratio of TWA trajectories falling into either stroboscopic basins is strongly influenced by the initial state's uncertainty width, which scales with $1/\sqrt{N}$. 
\subsection{Pre-thermal regime: phase flips}
\begin{figure}
    \centering
    \includegraphics[width=1.0\linewidth]{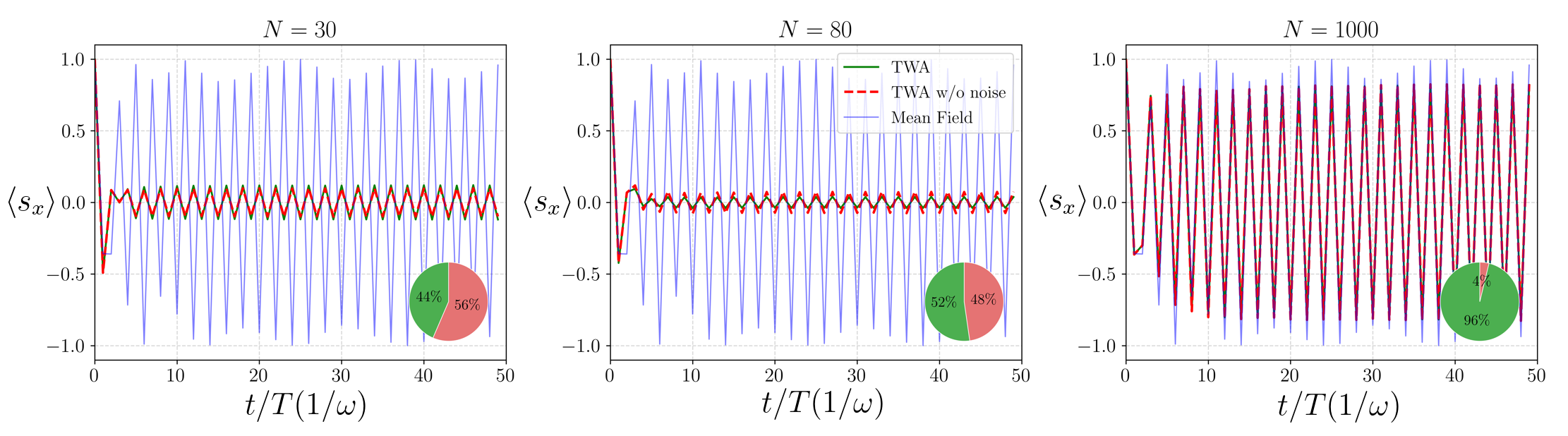}
    \caption{Stroboscopic dynamics of $\langle s_x\rangle$ in the pre-thermal regime computed by mean field (blue), TWA (green), and TWA without noise (dashed red) for $N=30,~80$, and $1000$. The pie charts show the percentages of TWA trajectories that are in phase (green) or out of phase (red) with the mean-field trajectory. Number of TWA trajectories: 10,000. Parameters: $\epsilon=0,~\kappa=0.05,~\lambda_0=1.0$}
    \label{fig:phase_vs_N_ep=0.00}
\end{figure}
The TWA average is a sum over trajectories that each converge to one of the two attractors. 
If a fraction $f_\text{in}$ of trajectories reaches the in-phase attractor, the stroboscopic amplitude of $\langle s_x \rangle$ is roughly $(f_\text{in}-f_\text{out})A_\text{MF}$, where $A_\text{MF}$ is the mean-field amplitude.
\begin{itemize}
    \item At small $N~(N=30)$, the initial cloud is wide and overlaps the blue region of Fig.~\ref{fig:mf_phase_space} substantially. 
    For the full TWA implementation, most trajectories (56\%) end on the out-of-phase attractor, so the net oscillation is out of phase with mean field, with $f_\text{in}-f_\text{out}=-0.12.$ (Fig.~\ref{fig:phase_vs_N_ep=0.00}).
    \item As $N$ increases, the cloud shrinks and fewer trajectories cross the basin boundary.
The out-of-phase majority is lost, and the net amplitude passes through a minimum: at $N=80$, $f_\text{in}-f_\text{out}=+0.04.$
    \item For $N \to \infty$, almost all trajectories follow the mean field ($f_\text{in}-f\text{out}=0.92$ at $N=1000$), and the amplitude approaches $A_\text{MF}$.
\end{itemize}

The DTC amplitude therefore first decreases and then increases with $N$, which is the non-monotonic scaling reported in the main text. 
The TWA curves with and without noise in Fig.~\ref{fig:phase_vs_N_ep=0.00} nearly coincide, so the quantum noise from cavity loss does not affect this trend.
The uncertainty in the initial state alone is responsible for this.
\subsection{Transient regime: chaotic decorrelation}
\begin{figure}
    \centering
    \includegraphics[width=\linewidth]{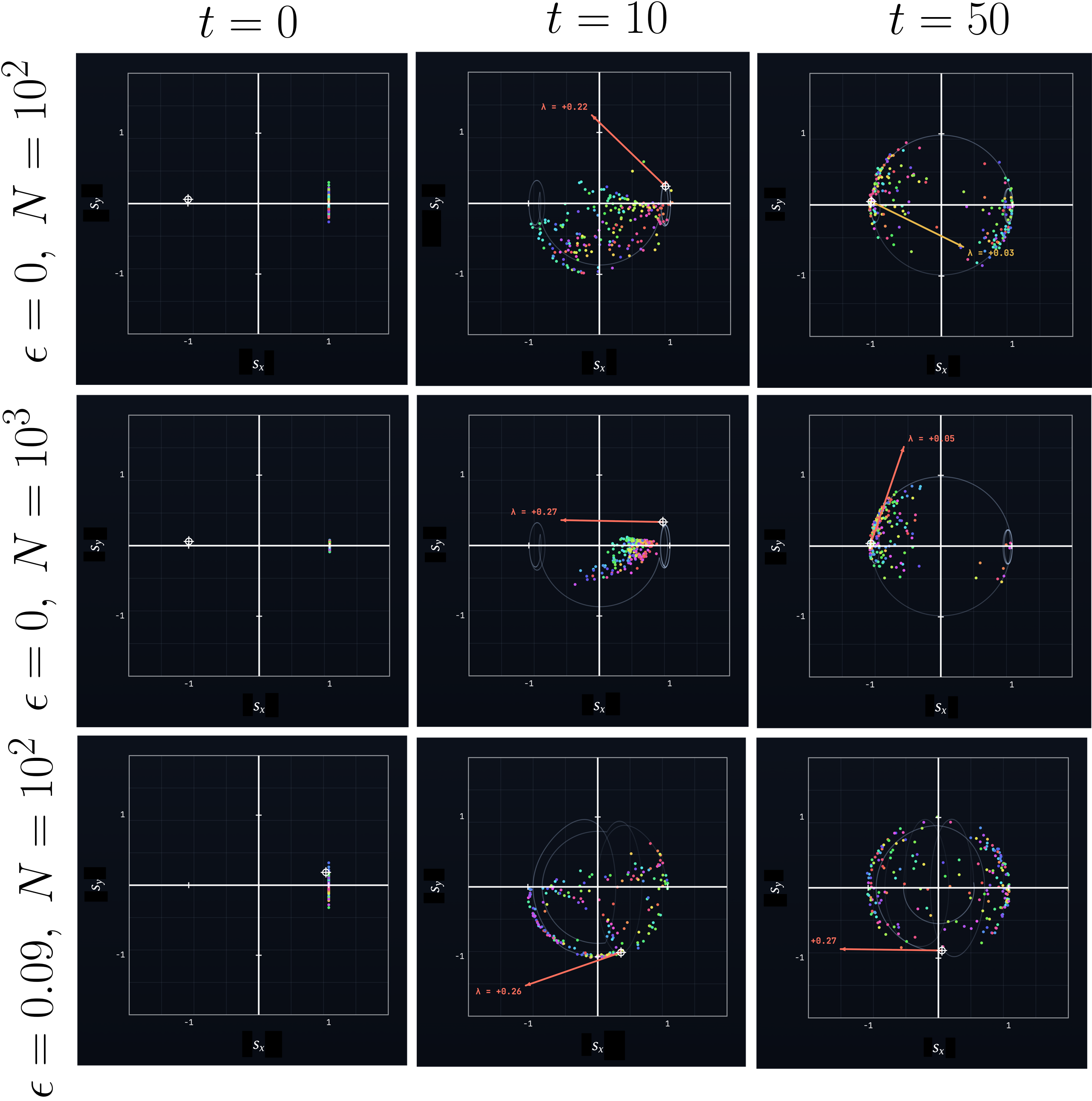}
    \caption{TWA trajectories in the $( s_x,s_y)$ plane at $t=0,10,$ and $50$ $\omega^{-1}$ for the pre-thermal regime ($\epsilon=0$; $N=10^2$ top row, $N=10^3$ middle row) and the transient regime ($\epsilon=0.09$, $N=10^2$, bottom row). The orange arrows represent the Lyapunov vectors associated with the largest, most positive Lyapunov exponents. Each colored dot represents a TWA trajectory. Parameters: $\kappa=0.05,~\lambda_0=1.0$.}
    \label{fig:lyapunov}
\end{figure}
The phase flips described above also occur in the transient regime, but here they are combined with a second effect.
Fig.~\ref{fig:lyapunov} illustrates this through the largest finite-time Lyapunov exponent $\Lambda$.

For $\epsilon=0$, the spin and cavity field rotate at similar frequencies ($\epsilon=(\omega-\omega_c)/2$). 
As a result, their mutual feedback is nearly the same in every period, and the trajectories settle quickly.
At $t=10$, $\Lambda$ is still clearly positive ($0.22$ for $N=100$ and $0.27$ for $N=1000$), indicating a chaotic transient.
By $t=50$, it has decayed to nearly zero ($0.03$ and $0.05$ respectively), consistent with convergence to a limit cycle, the period-doubled attractor.

For $\epsilon=0.09$, the frequency mismatch changes the relative phase of spin and cavity field from period to period and scrambles the trajectories. $\Lambda$ stays large at intermediate times ($0.26$ at $t=10$ and $0.27$ at $t=50$), and the trajectories spread around the whole circle in Fig.~\ref{fig:lyapunov} (bottom row).
\begin{figure}
    \centering
    \includegraphics[width=1.0\linewidth]{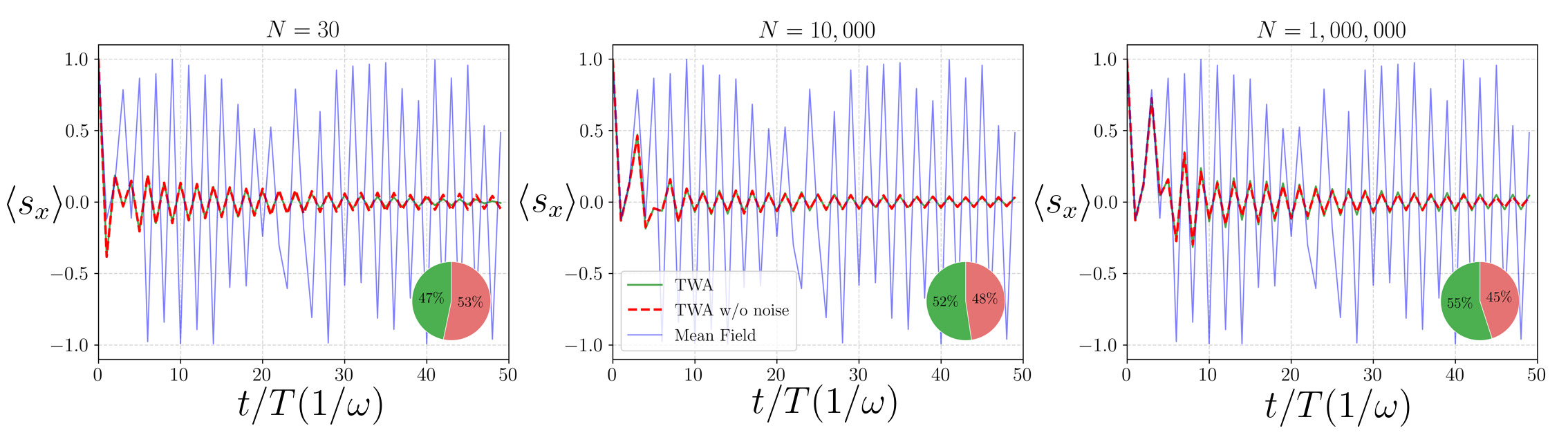}
    \caption{Stroboscopic dynamics of $\langle s_x\rangle$ in the transient regime computed by mean field (blue), TWA (green), and TWA without noise (dashed red) for $N=30,~10^4,$ and $10^6$. The pie charts show the percentages of TWA trajectories that are in phase (green) or out of phase (red) with the mean-field trajectory. Number of TWA trajectories: 10,000. Parameters: $\epsilon=0.09,~\kappa=0.05,~\lambda_0=1.0$}
    \label{fig:phase_vs_N_ep=0.09}
\end{figure}
Because the chaotic trajectories decorrelate, their contributions partially cancel in the average, and the averaged amplitude decays (Fig.~\ref{fig:phase_vs_N_ep=0.09}).
The in/out-of-phase fractions again cross from an out-of-phase majority at small $N$ to an in-phase majority at large $N$.
The approach to mean field is much slower than in the pre-thermal regime, however ($f_\text{in}=55\%$ even at $N=10^6$).
The combination of chaotic decorrelation and the $N$-dependent phase flip produces the non-monotonic $N$-dependence of the DTC decay rate in the transient regime.

\end{document}